\documentclass[prc,twocolumn,preprintnumbers,showpacs,superscriptaddress]{revtex4}
\usepackage[T1]{fontenc}
\usepackage{amsmath,aas_macros}
\usepackage{natbib}
\usepackage{soul}
\usepackage{xcolor}
\usepackage{lipsum}
\usepackage{float}
\usepackage{graphicx}	
\usepackage{multirow,makecell}
\usepackage{placeins}
\usepackage{bm,amsfonts,hhline,amssymb,microtype,float,latexsym,epsf,mathtools,times,makecell,array}
\usepackage{tabularx}
\usepackage{epstopdf}
\usepackage{gensymb}
\usepackage{pifont}
\usepackage{hyperref}
\usepackage{lineno}
\hypersetup{colorlinks=true, linkcolor=red,filecolor=blue,urlcolor=magenta, citecolor=blue}

\definecolor{ForestGreen}{HTML}{228B22}

\newcommand{\Qsat}{Q_{\text {sat }}} 
\newcommand{\Lsym}{L_{\text {sym }}} 
\newcommand{\OmKep}{\Omega_{\mathrm{Kep}}}
\newcommand{\fKep}{f_{\mathrm{Kep}}}
\newcommand{\bea}{\begin{eqnarray}}
\newcommand{\eea}{\end{eqnarray}}

\usepackage[normalem]{ulem}  

\definecolor{darkred}{rgb}{0.7,.1,.2}

\begin{document}
\title{Effect of symmetry energy on properties of rapidly rotating neutron stars and universal relations}

\author{Pion Sudarshan Yeasin}
\email{psyeasin@icloud.com$^{*}$}
\affiliation{Institute of Theoretical Physics, University of Wroc\l{}aw, 50-204 Wroc\l{}aw, Poland}

\author{Stefanos Tsiopelas}
\email{stefanos.tsiopelas2@uwr.edu.pl$^{*}$}
\affiliation{Institute of Theoretical Physics, University of Wroc\l{}aw, 50-204 Wroc\l{}aw, Poland}

\author{Armen Sedrakian}
\email{sedrakian@fias.uni-frankfurt.de$^{\ddagger}$}
\affiliation{Institute of Theoretical Physics, University of Wroc\l{}aw, 50-204 Wroc\l{}aw, Poland}
\affiliation{Frankfurt Institute for Advanced Studies, D-60438 Frankfurt am Main, Germany}

\author{Jia-Jie Li}
\email{jiajieli@swu.edu.cn$^{\star}$}
\affiliation{School of Physical Science and Technology, 
             Southwest University, Chongqing 400715, China}    

\date{September 1, 2025}

\begin{abstract}   
We investigated universal relations for compact stars rotating at
  the Keplerian (mass-shedding) limit, which is highly relevant for
  understanding the rapidly rotating objects formed in the aftermath
  of a neutron star-neutron star merger.  Our analysis is based on a
  set of nucleonic equations of state (EoSs) featuring systematic
  variations in the symmetry energy slope parameter $\Lsym$ and the
  isoscalar skewness parameter $\Qsat$, varied within ranges that are
  broadly consistent with current laboratory and astrophysical
  constraints. The global observable properties of isolated maximally
  rotating stars are examined, focusing on the mass-radius relation,
  moment of inertia, quadrupole moment, and the Keplerian (maximum)
  rotation frequency, as well as their variations in the
  $\Lsym$-$\Qsat$ parameter space.  Next, we demonstrate that, in the
  limit of Keplerian rotation, universal relations remain valid across
  the same set of EoSs characterized by varying $\Lsym$ and
  $\Qsat$. In particular, we present explicit results for the moment
  of inertia ($I$) and quadrupole moment ($Q$) as functions of
  compactness, as well as for the moment of inertia-quadrupole moment
  relation. All of these relations exhibit excellent universality,
  with deviations typically within a range from a few percent to 10\%
  across a wide range of parameters. Additionally, we verify for our
  set of EoSs that the universality of $I$-$Q$ holds to higher accuracy
  (at the level of 1\%) in the slow-rotation approximation compared with
  the Kepler limit, where the relative error increases up to $\lesssim
  10\%$.  Our findings support the applicability of $I$-Love-$Q$-type
  universal relations in observational modeling of maximally rotating
  compact stars and the gravitational wave emitted by them, when
  accounting for significant variation in the symmetry energy and
  high-density behavior of the nuclear EoS.
\end{abstract}

\maketitle

\section{Introduction}
\label{sec:intro}
In recent years, the equation of state (EoS) of cold compact stars (CSs) has been significantly constrained due to a range of new astrophysical observations, including measurements of stellar masses, radii, and tidal deformability~\cite{LVC:2017,Demorest:2010bx,Cromartie:2019kug,Fonseca:2021}. These constraints have been further reinforced by experimental data from nuclear physics and advances in {\it ab initio} calculations of pure neutron matter~\cite{Lattimer:2023,Drischler:2019}.  Covariant density functionals (CDFs) offer a fast and reliable approach to integrate physical constraints derived from both the bulk properties of nuclear systems and astrophysical observations, starting from baryon-meson Lagrangians~\cite{Oertel_RMP_2017,Sedrakian:2023}. They enable access to microscopic details, including self-energies (classified according to their Lorentz structure), matter composition, chemical potentials, and effective masses. Furthermore, CDFs allow for flexible adjustments of model parameters in response to new information and evolving constraints on the EoS or other aspects of both micro-- and macrophysics.

Recently, Li and Sedrakian~\cite{Li2023ApJ} (hereafter LS23) developed three families of CDFs based on the DDME2, DD2, and MPE parametrizations~\cite{Lalazissis:2005,Typel:2018}, introducing variations in the slope of the symmetry energy and the skewness of symmetric matter at saturation density. (The tables of EoSs are publicly available at ~\cite{armen_sedrakian_2023_8350680}). These newly constructed CDFs facilitate systematic studies of the sensitivity of various astrophysical scenarios to changes in these parameters. An effective way to leverage these results is to adapt existing codes that utilize the original CDFs, such as DDME2, DD2, and MPE~\cite{Typel:2018}, to extract astrophysical observables. These families are designed to span a broad region of the mass-radius diagram for CSs that are consistent with current astrophysical constraints. Such comprehensive coverage enables systematic investigations into how astrophysical simulations depend on key yet poorly determined parameters of the nuclear EoS, such as the slope of the symmetry energy $\Lsym$ and the skewness of symmetric matter $\Qsat$ at nuclear saturation density. The integral parameters of CSs in the static limit were computed in Ref.~\cite{Li2023ApJ}.  Furthermore, their finite-temperature counterparts are also available in the CompOSE format~\cite{Tsiopelas2024EPJA}, which are suitable for simulations of supernovas and binary neutron star mergers. 

One of the objectives of this work is to study rapidly rotating CSs for the cold EoS models of LS23. We aim to extract information on various parameters, such as the mass, radius, moment of inertia, etc., at the Keplerian frequency (the mass-shedding limit), which represents the maximum rotation rate compatible with equilibrium. It should be noted that the fastest known pulsar J1748-2446ad rotates at a spin frequency 716 Hz
(its period is 1.396 ms) which is well below the typical frequency $\fKep\sim 1200-1500 \mathrm{~Hz}$ for a canonical $1.4~M_{\odot}$ mass neutron star. However, the understanding of maximally rapidly rotating CSs has attracted renewed interest in the context of neutron star mergers and the GW170817 event. It has been argued that this event places an upper limit on the maximum mass of a {\it static} CS using the scenario in which the merger event results in the formation of a supramassive compact object, which eventually collapses to a black hole~\cite{Rezzolla2018ApJ,Shibata2019PhRvD,Khadkikar2021PhRvC}. To find the bound on the static mass, the universal relations that relate the masses of Keplerian configurations to their nonrotating counterparts were used.

Motivated by this argument, our second aim is to examine the universal relations among the global properties of CSs that rotate rigidly. The universal relations among various global parameters of CSs have been known for a long time, but became popular after the observation that the moment of inertia ($I$), tidal deformability ($\Lambda$), and quadrupole moment $(Q)$ obey universality; see Refs.~\cite{Yagi:2013a,Yagi:2013b}. Universal relations have been extensively studied in various contexts, which can be roughly categorized as follows (note however, that some works belong to multiple categories): (a) extraction of physical information from gravitational-wave data and the static or slow-rotation limits~\cite{Maselli_PRD_2013,Baubock:2013,Steiner:2015,Silva:2017,Weijb:2018,Lim:2018,Kumar:2019,Wendh:2019,Jiangnan:2020,Sunwj:2020,Godzieba:2020,Nedora:2021,Saes:2021,Kuan:2022,Wangsb:2022,Zhaotq:2022,Manohara2024}; (b) rapidly rotating CSs~\cite{Doneva:2013,Pappas:2014,Chakrabarti:2014,Cipolletta:2015,Breu:2016,Riahi:2019, Raduta_MNRAS_2020,Koliogiannis:2020, Khadkikar:2021,Konstantinou:2022,Largani:2022,Li2023PhRvC}; (c) hot proto-CSs and postmerger remnants~\cite{Martinon:2014,Marques:2017,Lenka:2018,Stone:2019,Khadkikar:2021}; (d) magnetized CSs~\cite{Haskell:2013}; (e) CSs containing heavy baryons~\cite{Raduta_MNRAS_2020,Li2020PhLB,Stone:2019,Li2023PhRvC, Khadkikar:2021}; (f) CSs with quark matter phases~\cite{Yagi:2013a,Yagi:2014,Weijb:2018, Steiner:2015,Paschalidis:2017,Annala:2021,Largani:2022}; (g) CSs in alternative theories of gravity~\cite{Staykov:2016,Doneva2018,Lam_2025PhRvL}; and (h) stars in binary systems~\cite{Yagi:2016,Doneva:2016,Danchev:2021,Manoharan2021}.
  Universality in this context implies that these relations, well established in the static limit, are largely independent of the underlying EoS.   If established, the universality is particularly valuable for interpreting observational data, as it helps to reduce uncertainties associated with the EoS. Given their significance, it is worthwhile to explore these relations for the class of EoSs introduced in LS23, which features systematic variations in the symmetry energy slope and the high-density behavior of matter, independent of isospin effects.

  This paper is organized as follows. In Sec.~\ref{sec:Kepler} we discuss the properties of CSs with varying
  slope of symmetry energy $L_{\text {sym }}$  and skewness $Q_{\text {sat }}$ [defined below in Eq.~\eqref{eq:CDF}], focusing on the scaling of various integral quantities of the stars on these input parameters. In Sec.~\ref{sec:Universal} we present the evidence for the universality of the relations between the integral parameters of the star on the input EoSs with continuous variations of $L_{\text {sym }}$ and $Q_{\text {sat }}$ parameters. Our conclusions are collected in Sec.~\ref{sec:Conclusions}. All computations are done with the public domain RNS code (https://github.com/cgca/rns).
 Geometric units $G=c=1$ are used throughout the manuscript unless otherwise noted. 

\section{Rapidly rotating stars}
\label{sec:Kepler}

To set the stage, we start with the definition of the nuclear characteristic parameters, which
appear in the  well-known Taylor expansion of the cold nuclear matter energy density
around the saturation density,
\bea\label{eq:CDF}
E(\chi, \delta) &\simeq& E_{\text{sat}} + \frac{1}{2}K_{\text{sat}}\chi^2
+ \frac{1}{6}Q_{\text{sat}}\chi^3 
+ E_{\text{sym}}\delta^2 \nonumber\\
&+& L_{\text{sym}}\delta^2\chi
+ {\mathcal O}(\chi^4, \chi^2\delta^2),
\eea
where $\chi = (\rho - \rho_{\text{sat}})/(3\rho_{\text{sat}})$ and $\delta = (\rho_n - \rho_p)/\rho$, with $\rho_n$ and $\rho_p$ denoting the neutron and proton densities, respectively. The expansion coefficients characterize bulk nuclear matter: the incompressibility $K_{\text{sat}}$, skewness $Q_{\text{sat}}$, 
symmetry energy $E_{\text{sym}}$, and slope of the symmetry energy $L_{\text{sym}}$. LS23 used paramterizations of CDFs with $E_{\text{sat}} = -16.14$ MeV, $K_{\text{sat}} = 251.15$ MeV, $E_{\text{sym}} = 32.31$ MeV, and $\rho_{\text{sat}} = 0.152$ fm$^{-3}$, which are the same as in  the DDME2 parametrization~\cite{Lalazissis:2005}.  With these baseline values fixed, the LS23 parametrizations of CDFs vary the less well-constrained higher-order nuclear matter characteristics, specifically the symmetry energy slope $L_{\text{sym}} $ and skewness $Q_{\text{sat}}$.
\begin{figure*}[t] \centering \includegraphics[width=\columnwidth]{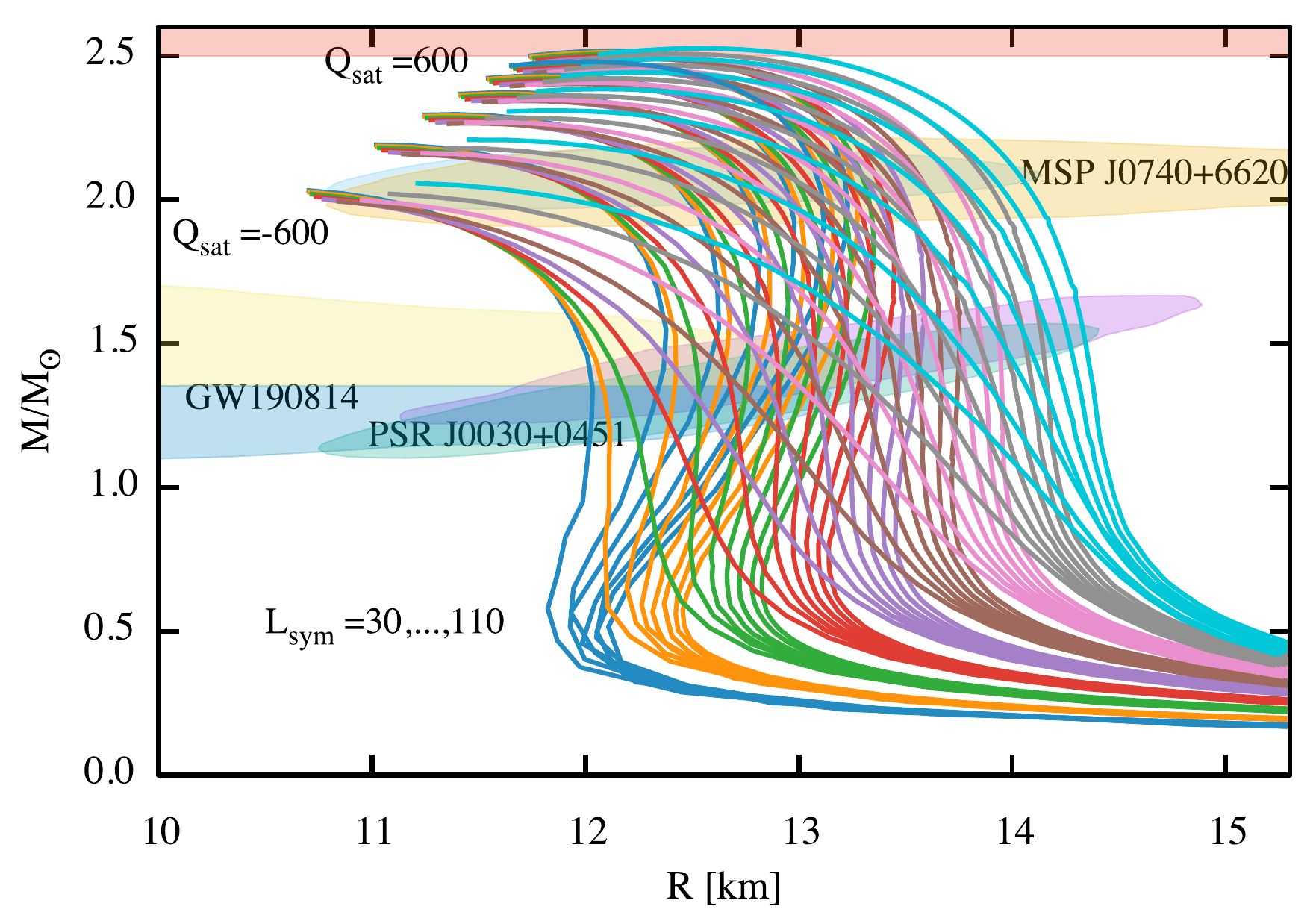} \includegraphics[width=\columnwidth]{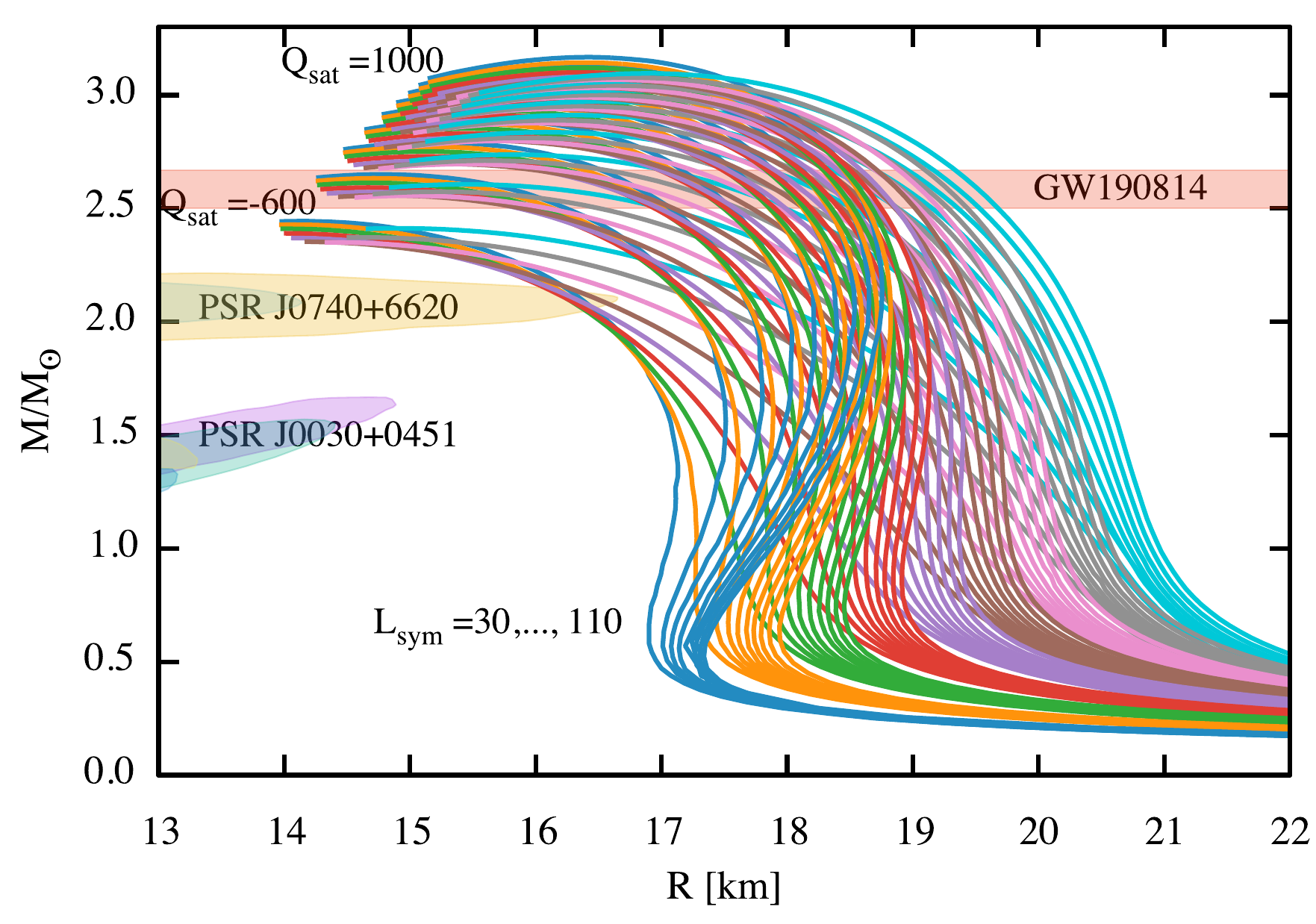} \caption{Left panel: mass-radius relations for nucleonic EoS models of nonrotating CSs with different pairs of values of $\Qsat$ and $\Lsym =30$ (blue), 40 (orange), 50 (green), 60 (red), 70 (violet), 80 (brown), 90 (magenta), 100 (grey), and 110 (cyan) [MeV].
 The color regions show the 90\% confidence interval (CI) ellipses from each of the two NICER modeling groups for PSR J0030+0451 and J0740+6620~\cite{NICER:2021a,NICER:2021b}, the 90\% CI regions for each of the two CSs that merged in the gravitational-wave event GW170817~\cite{Abbott_2019}, and finally the 90\% CI for the mass of the secondary component of GW190814~\cite{Abbott_2020}.  Right panel: same as in the left panel, but for CSs rotating at Keplerian frequency. The stellar radius corresponds to the equatorial radius.}
\label{fig:MR}
\end{figure*}

Figure~\ref{fig:MR} shows the mass-radius diagram for non-rotating CSs, obtained by solving the Tolman-Oppenheimer-Volkoff (TOV) equations, see also Fig.~3 of LS23, in comparison with the current observational constraints. The figure shows a collection of 81 different EoSs for each fixed set of low-order characteristics of DDME2, generated by varying the parameters within the ranges of $-600 \leq Q_{\text {sat }} \leq 1000$~MeV and $30 \leq$ $L_{\text {sym }} \leq 110$ MeV, with step sizes of $\Delta Q_{\text {sat }}=200$ MeV and $\Delta L_{\text {sym }}=10$ MeV. These parameter ranges are chosen as follows. At $Q_{\text{sat}}=-600$~MeV, regardless of the value of $L_{\text{sym}}$, the maximum mass $M_{\rm max}^{\rm TOV}$ reaches approximately $2.0 M_{\odot}$, consistent with the mass measurement of PSR J0740+6620~\cite{Cromartie:2019kug,Fonseca:2021}. For $Q_{\text{sat}} \geq 600$ MeV, the maximum mass of a CS is $M_{\rm max}^{\rm TOV} \gtrsim 2.5\,M_{\odot}$, matching the value inferred for the lighter companion of GW190814, $2.59_{-0.09}^{+0.08}\,M_{\odot}$~\cite{Abbott_2020}, if interpreted as a neutron star. The selected lower and upper bounds for $L_{\text{sym}}$ are consistent with the predictions of most density functionals and agree with the conservative experimental ranges currently under discussion (for reviews, see Refs.~\cite{Lattimer:2023,Sedrakian:2023}). 

Turning now to the maximally fast rotating CSs, we show in the right panel of Fig.~\ref{fig:MR} the gravitational masses of the stars as a function of equatorial circumferential radius for a maximally rotating star. For intermediate rotation frequencies, the radius lies between the Keplerian radius and the static radius shown in the left panel of  Fig.~\ref{fig:MR}. CSs that are born with rapid rotation evolve adiabatically, losing energy primarily through gravitational-wave emission and, to leading order, electromagnetic dipole radiation. This evolution can be effectively modeled by sequences of stellar configurations with constant rest mass and varying spin~\cite{Cook:1994,Stergioulas:1994,stergioulas2003rotating}. The most massive of these rapidly rotating stars may lack a stable nonrotating (static) counterpart, meaning that their spindown can ultimately lead to gravitational collapse into a black hole. Such stars are referred to as supramassive configurations. Since the constant baryon mass sequences are nearly parallel to the $R$ axis~\cite{Cook:1994}, it is seen that the most massive rotating configurations with masses $M\ge 2.5M_{\odot}$ are supramassive irrespective of the values of $\Lsym$ and $\Qsat$. Otherwise, the ordering of the stellar sequences with respect to the variations of these two parameters is analogous to the one seen in Fig.~\ref{fig:MR}, with a clear positive correlation between $\Lsym$ and the radius and $\Qsat$ and the maximum mass.

\begin{figure*}[t]
\centering
\includegraphics[width=\columnwidth]{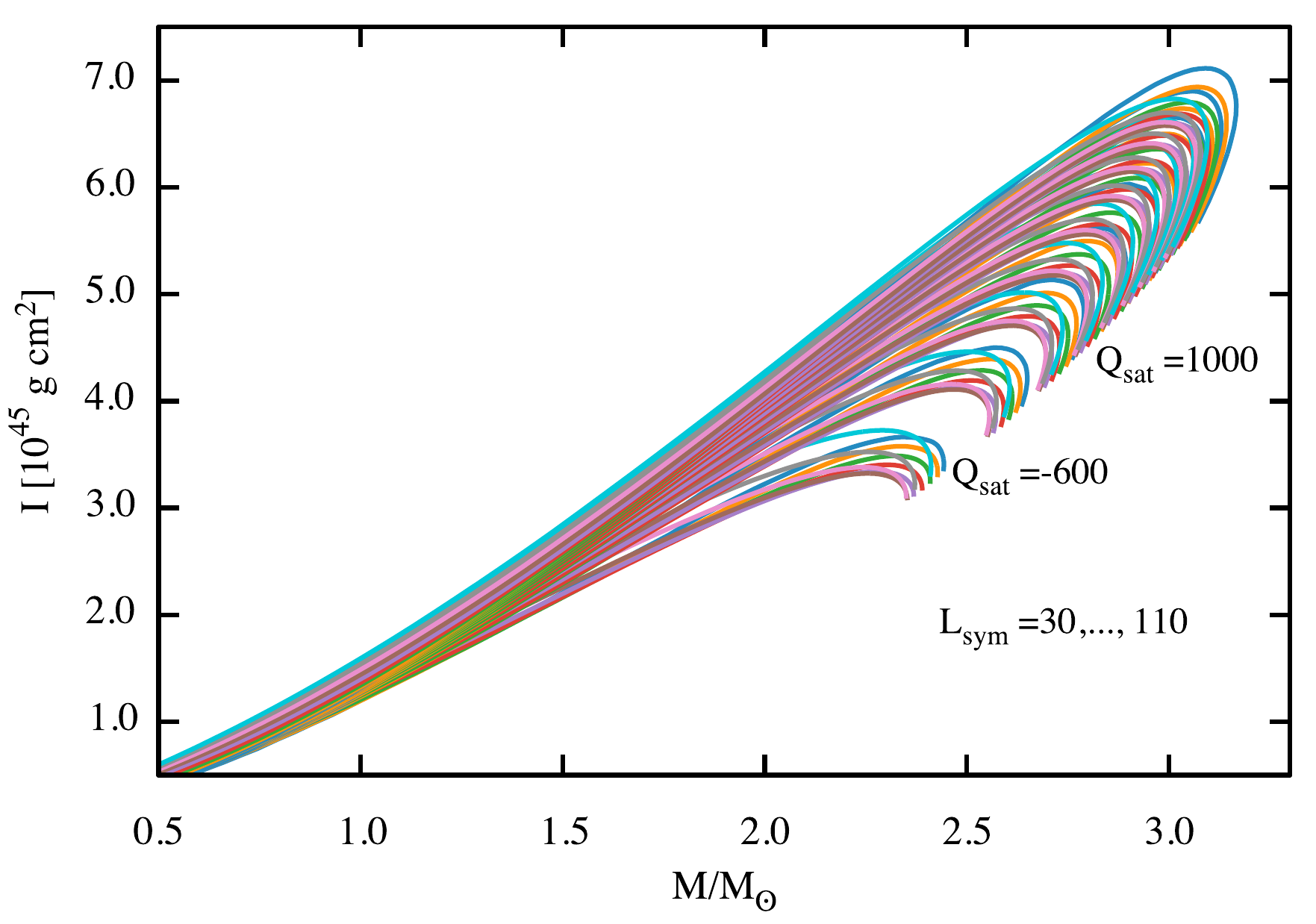}
\includegraphics[width=\columnwidth]{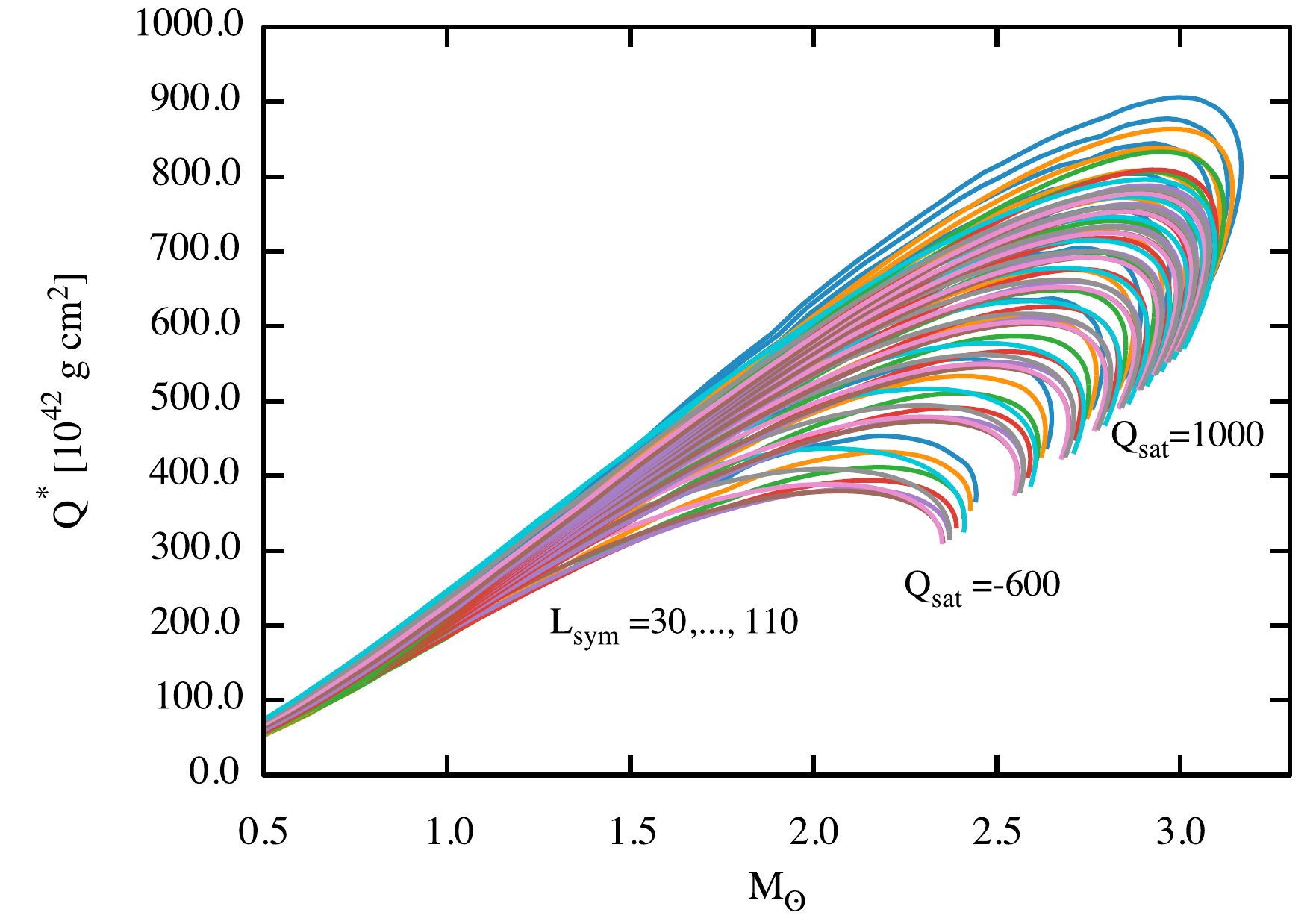}
\caption{Left panel: dependence of the moment of inertia of Keplerian CSs on  mass for varying values of $\Qsat$ and $\Lsym$ (in units of MeV) as indicated in the plot.  Right panel: dependence of the (dimensional)
  quadrupole moment of Keplerian CSs on mass for varying values of $\Qsat$ and $\Lsym$ as indicated in the plot.  The color convention is the same as in Fig.~\ref{fig:MR}.}
\label{fig:IQM}
\end{figure*}
For rotating CSs,  $I \propto MR^2$, where  $M$ and $R$ are the mass and radius of the star, respectively. As a result, models with larger  $\Qsat$, which typically support more massive configurations due to increased stiffness of the high-density EoS,  exhibit higher moments of inertia when other parameters are held fixed. Additionally, higher values of $\Lsym$, which characterizes the density dependence of the symmetry energy, are associated with stiffer EoSs at densities close to saturation. This leads to larger stellar radii for a given mass, thereby enhancing the moment of inertia due to its strong $R$ dependence. Consequently, stars with larger radii, favored by larger $\Lsym$, tend to have significantly higher moments of inertia at fixed mass and rotational frequency. These dependencies show the sensitivity of $I$ to key parameters of nuclear matter and its importance as a probe of the dense matter EoS through pulsar timing and gravitational-wave observations.

The left panel of Fig.~\ref{fig:IQM},  illustrates how the moments of inertia of Keplerian stellar models vary with stellar mass for selected values of $\Qsat$ and $\Lsym$, consistent with the scaling for slowly rotating stars. (We do not show the same relation in the static limit as it is analogous to the Keplerian limit, but with shifted scales). The right panel of Fig.~\ref{fig:IQM} illustrates the dependence of the
mass quadrupole moment of Keplerian stellar models on the stellar
mass, for selected values of $\Qsat$ and $\Lsym$. (We have checked
that the arrangement of the curves is the same in the static limit).
The quadrupole moment quantifies the rotational deformation of the
star. In the slow-rotation and/or Newtonian approximation it scales as
$Q \propto k R^5 \Omega^2 / G$ where $k$ is a structural constant; see
Eq. (70) in \cite{Yagi:2013b}. In the same approximation, the Kepler
frequency scales as $\Omega_{\rm Kep} \propto GM/R^3$, a scaling that
remains valid for highly relativistic CSs; see
Eq.~\eqref{eq:OmKep_scaling} below. Combining these scalings, we find
for the quadrupole moment of Keplerian stars $Q_{\rm Kep}\propto
kMR^2$, which is the same in geometric units.  The $M(R)$ function for
Keplerian CSs (Fig.~\ref{fig:MR}, right panel) varies weakly with the
radius for masses $M\ge M_{\odot}$ not too close to the maximum,
implying that $Q_{\rm Kep}$ scales linearly with mass; close to the
maximum, the mass stays nearly constant, while the radius decreases
abruptly, thus also decreasing $Q_{\rm Kep}$ according to the scaling
above.  According to Fig.~\ref{fig:IQM} models with larger $\Qsat$
values, which allow for more massive Keplerian configurations, tend to
exhibit higher quadrupole moments when other parameters are held
fixed. Furthermore, stars with larger radii, associated with higher
values of $\Lsym$, also display larger quadrupole moments.  This is
consistent with the approximate scaling above.  Hence, quadrupolar
deformation is a sensitive probe of variations in the nuclear matter
parameters $\Qsat$ and $\Lsym$, which govern the stiffness of the EoS
and the resulting stellar structure.
\begin{figure*}[t]
\centering
\includegraphics[width=\columnwidth]{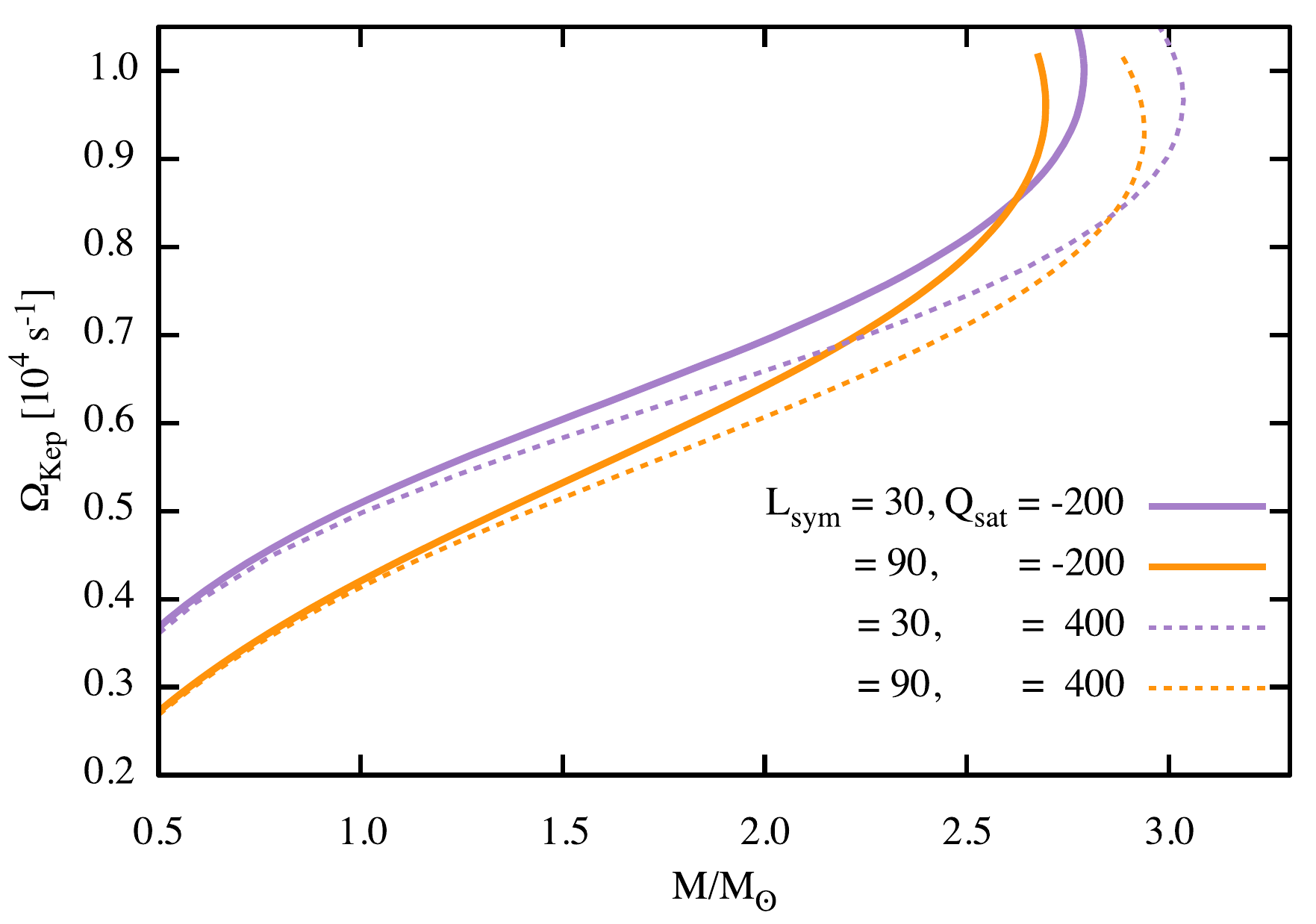}
\includegraphics[width=\columnwidth]{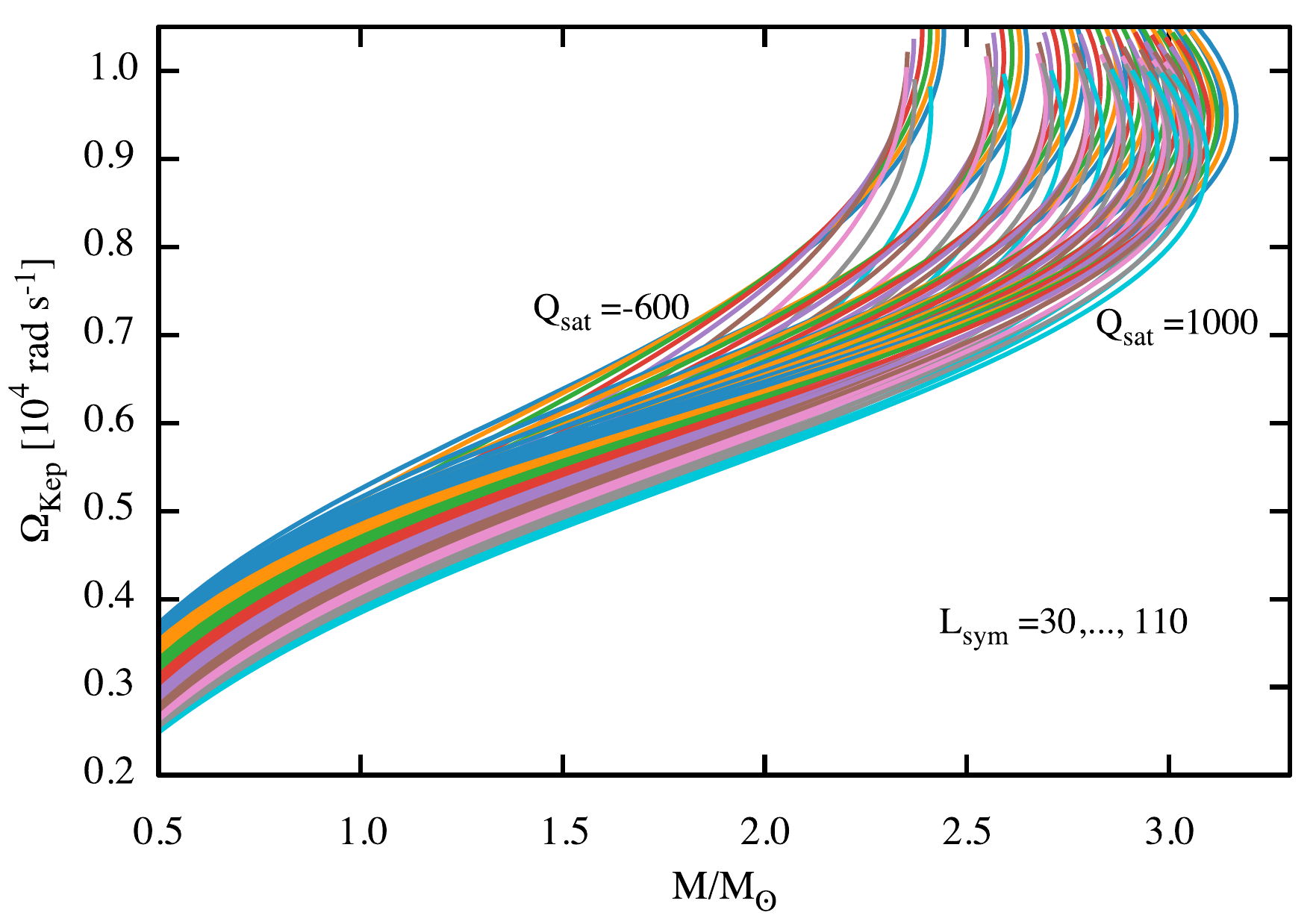}
\caption{Left panel: dependence of the Keplerian frequency on the mass of configuration for a pair of values of $\Lsym = 30$ (violet) and 90 (orange) and $\Qsat$ (in units of MeV)  as indicated in the plot. Right panel: same as in the left panel, but for the full set of EoSs. The color convention is the same as in Fig.~\ref{fig:MR}.
}
\label{fig:OmM}
\end{figure*}
\begin{figure*}[t]
\centering
\includegraphics[width=\columnwidth]{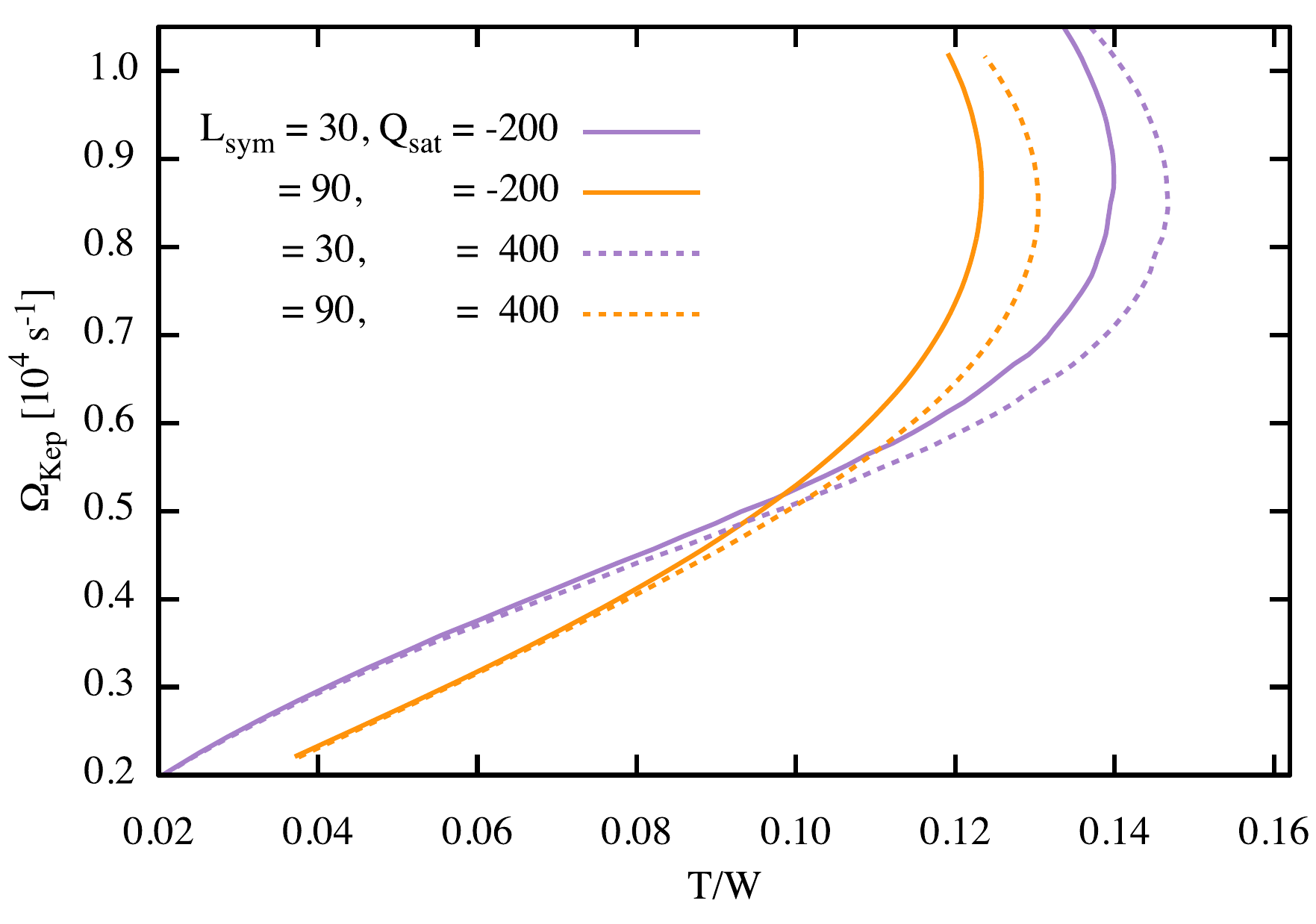}
\includegraphics[width=\columnwidth]{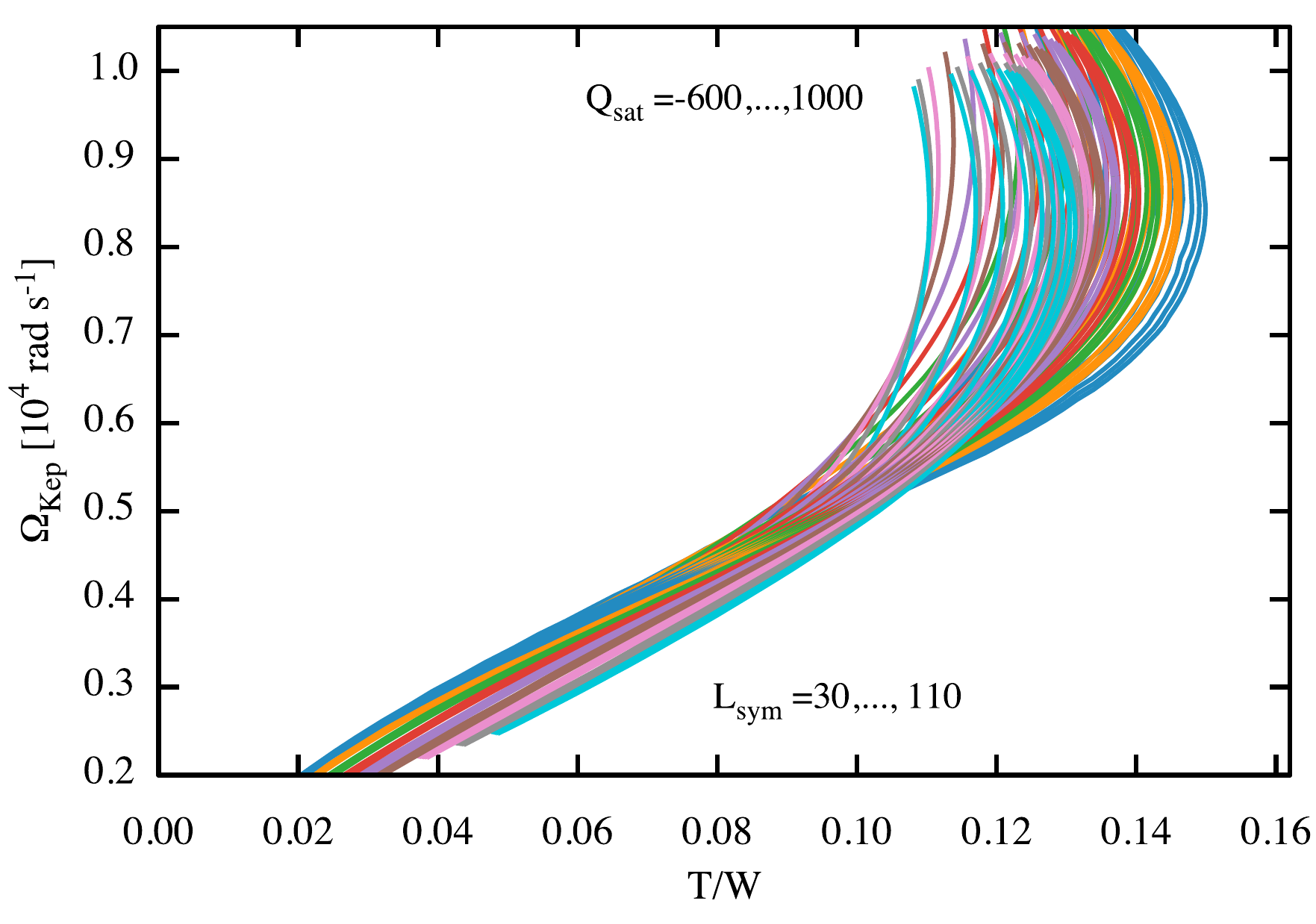}
\caption{Left panel: dependence of Keplerian frequency $\OmKep$ on  kinetic-to-gravitational energy ratio $T/W$ for a pair of values of $\Lsym = 30$ (violet) and 90 (orange) and $\Qsat$ (in units of MeV)  as indicated in the plot. Right panel: same as in the left panel, but for the full set of EoSs. The color convention is the same as in Fig.~\ref{fig:MR}.
  }
\label{fig:OmTW}
\end{figure*}

Next, recall the well-known relation between the Keplerian angular frequency and the nonrotating star's mass and radius~\cite{Cook:1994,Haensel2009,Riahi:2019,Li2023PhRvC}
\bea\label{eq:OmKep_scaling}
\OmKep = 2\pi \fKep \approx 2\pi  f_0
\left(\frac{M}{M_{\odot}}\right)^{1 / 2}\left(\frac{R}{10 \mathrm{~km}}\right)^{-3 / 2}~\rm{kHz} \eea
where $f_0 = 1.04$-$1.08$.

Figure~\ref{fig:OmM} shows the dependence of the Keplerian frequency on
mass for an illustrative choice of four EoSs  (left panel) and the full set of EoSs (right panel).
The general structure of $\fKep(M)$ can be understood from the scaling in Eq.~\eqref{eq:OmKep_scaling} with the mass and radius: in the low- to intermediate-mass range (but for $M\ge 0.5M_{\odot}$), the radius remains nearly constant while the mass increases, leading to a gradual rise in $\fKep$. In contrast, in the high-mass regime, the mass varies little but the radius decreases sharply. As a result, $f_{\text {Kep }}$ increases rapidly, in accordance with Eq.~\eqref{eq:OmKep_scaling}.  The left panel of Fig.~\ref{fig:OmM} shows that, for fixed $\Qsat$ and mass, a smaller value of $\Lsym$ leads to a higher $\fKep$ for masses not too close to the maximum. This trend is consistent with the fact that a smaller $\Lsym$ implies a smaller radius, which, according to Eq.\eqref{eq:OmKep_scaling}, results in a higher $\fKep$.
In the high-mass limit, the curves converge because the behavior is primarily governed by the common value of 
$\Qsat$. As the mass approaches the maximum, the curves cross and their ordering reverses: a smaller
$\Lsym$ --corresponding to a softer EoS--supports a higher maximum mass, effectively stretching the associated curve
further to the right. The same panel also shows that for the same $\Lsym$ and mass, a smaller $\Qsat$ implies a large $\fKep$. Thus, we conclude the following:
\begin{itemize}
\item When all other EoS parameters are held fixed and the mass is constant, a larger symmetry energy slope 
$\Lsym$ leads to a lower Keplerian frequency  $\fKep$.
\item Similarly, for fixed mass and fixed EoS parameters, a smaller skewness parameter 
$\Qsat$ results in a higher  $\fKep$.
\end{itemize}
Figure~\ref{fig:OmTW} shows the dependence of $\fKep$ on the kinetic-to-gravitational energy ratio $T/W$ for an illustrative choice of four EoSs (left panel) and the full set of EoSs (right panel). This dependence is useful for locating the dynamical bar-mode instability for $T/W\ge 0.27$ and secular (dissipation-driven) bar-mode instability for $T/W\ge 0.14$ for uniformly rotating CSs.  The Newtonian results for $T/W$ do not provide useful scaling for our case.  For example, for incompressible Maclaurin spheriods, $T / W= {\Omega^2 R^3}/{5G M}$; see Chap. 4 of Ref.~\cite{Chandrasekhar1969}. Substituting Eq.~\eqref{eq:OmKep_scaling} here shows that,
in the Newtonian limit, there is no explicit dependence of 
$T/W$ on mass or radius. Therefore, we rely below on the numerical results shown in the left panel of Fig.~\ref{fig:OmTW} to draw the following conclusions:
\begin{itemize}
\item When all other EoS parameters are held fixed and $T / W$ is constant, a smaller symmetry energy slope $\Lsym$ leads to a higher Keplerian frequency
  $\fKep$ in the low-$T / W \lesssim 0.1$ regime. However, this trend reverses at higher $T / W$, near and beyond the secular instability limit of 0.14.
\item Similarly, for fixed $T / W$ and fixed EoS parameters, a smaller skewness parameter $\Qsat$ results in a higher Keplerian frequency $\fKep$ across the entire $T / W$ range.
\end{itemize}  
The trends stated above are also seen in the right panels of Figs.~\ref{fig:OmM} and ~\ref{fig:OmTW} for the full collection of the EoS, which demonstrates the range of the possible value of the Kepler frequency for a broad range of the pair of parameters $30\le \Lsym\le 110$ MeV and $-600\le \Qsat\le 1000$ MeV.

\section{Universal relations}
\label{sec:Universal}
\begin{figure*}[bht]
\includegraphics[width=0.45\linewidth]{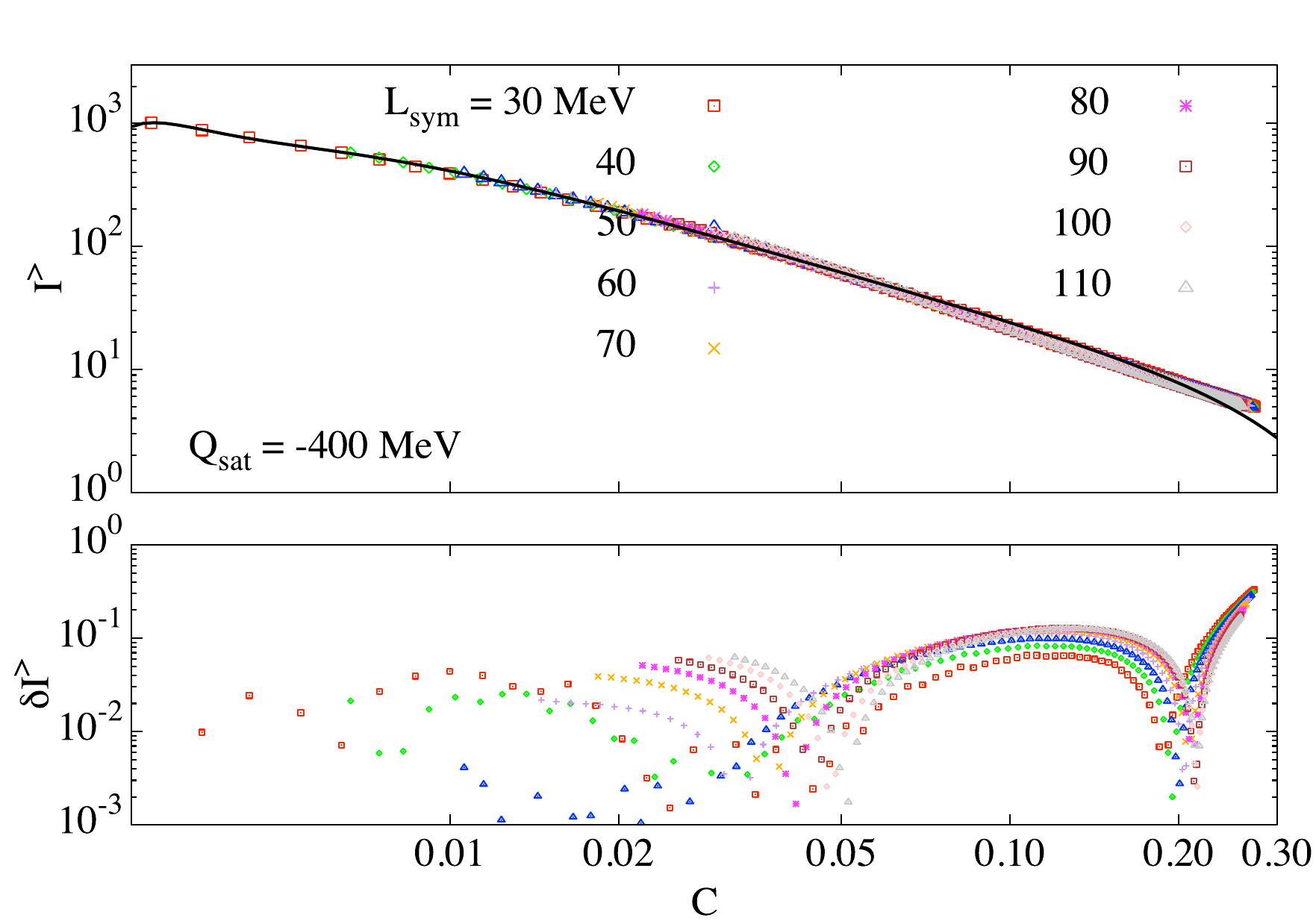}
\includegraphics[width=0.45\linewidth]{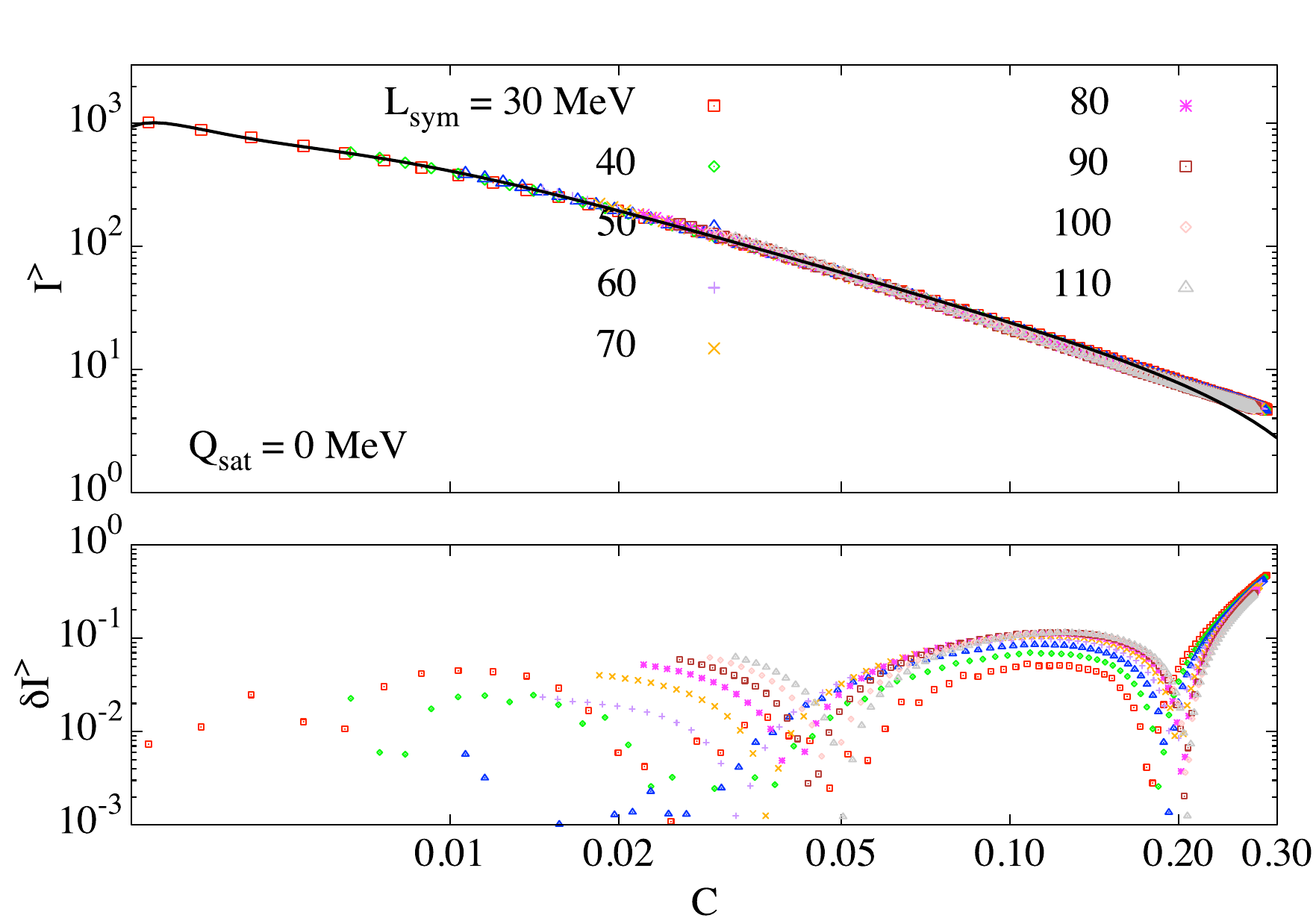}
\includegraphics[width=0.45\linewidth]{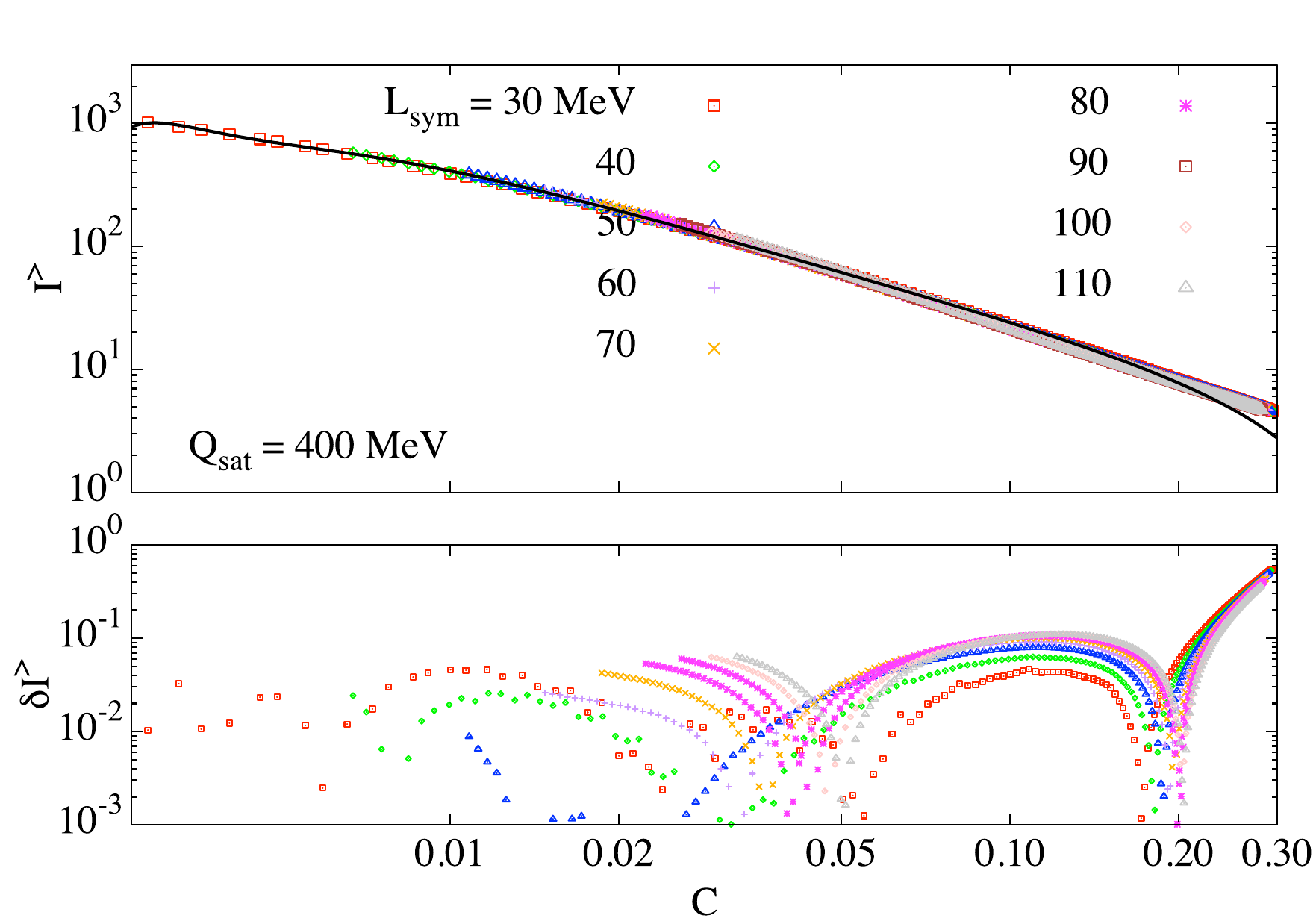}
\includegraphics[width=0.45\linewidth]{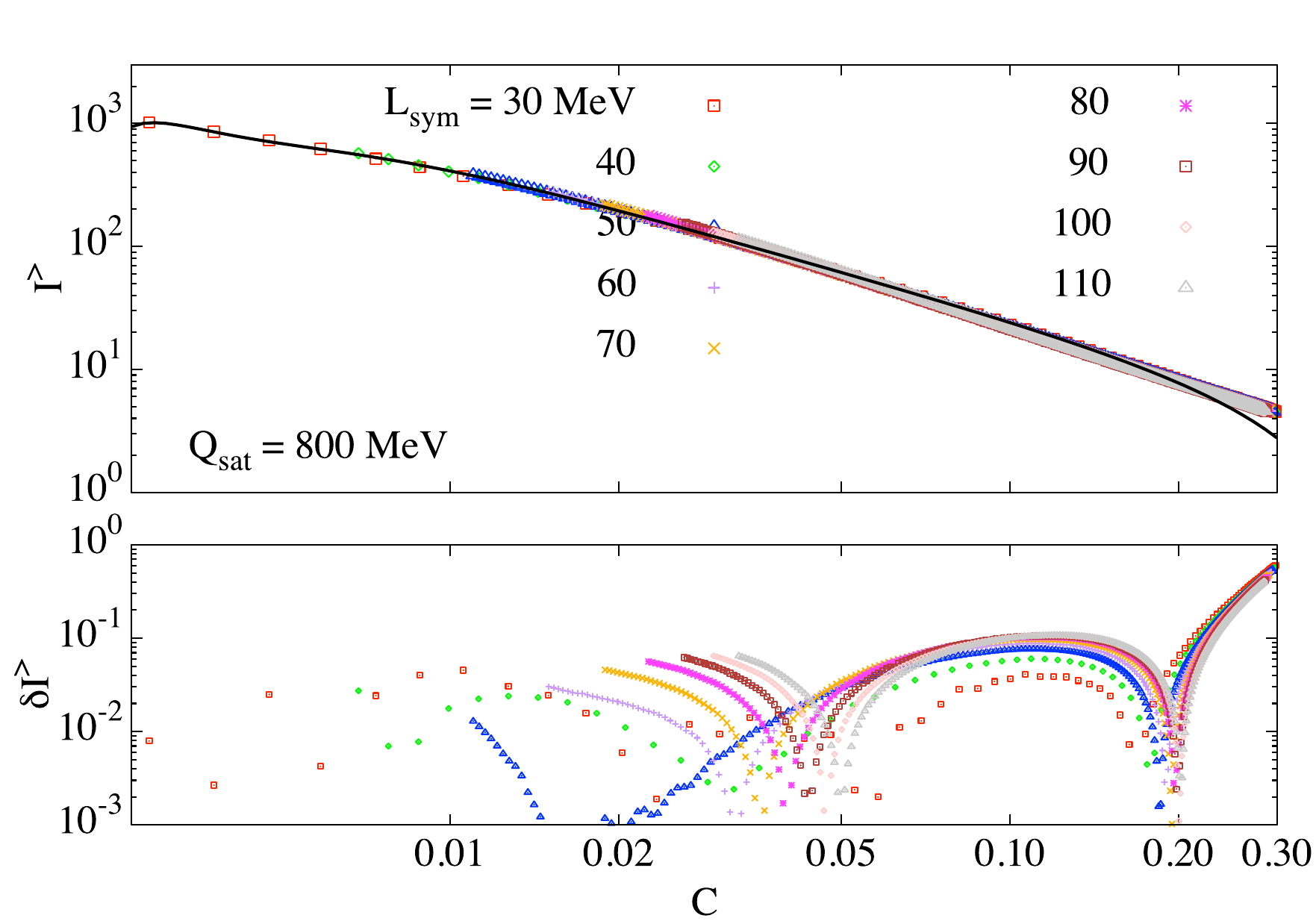}
\caption{Log-log plot of the dimensionless moment of inertia $I^{>}$ versus
  compactness $C$ for maximally fast rotating (Keplerian) configurations. The compactness is defined via the equatorial radius. It is limited from above by the maximum mass configuration (the curves are extended slightly in the unstable region where masses are below the maximum). The compactness is limited from below by the configuration with mass 0.1~$M_{\odot}$ for $L=30$~MeV and 0.5~$M_{\odot}$ for $L=110$~MeV. The bottom   panel shows the relative residual errors    $\delta I^{>}$, where 
  $\delta I^{>}= \vert ({I}^{>}-{I}^>_{\rm fit})\vert/{I}_{\rm fit}^{>}$
  with respect to the fit by Eq.~\eqref{eq:bar_I}.
  In each subpanel, the upper panel shows the data points by varying the
  parameter $\Lsym = 30$ (red $\square$), 40 (green $\lozenge$), 50 (blue $\triangle$), 60 (violet $+$), 70 (orange $\times$), 80 (magneta $\ast$), 90 (brown $\square$), 100 (rose $\lozenge$) and 110 (grey $\triangle$) in MeV for fixed $\Qsat$ and the best-fit curve based on a third-order polynomial in $C$. The lower panel displays the relative deviation between the numerical data and the fit.  }
\label{fig:I_bar_C_Fit_Error}
\end{figure*}
\begin{figure*}[hbt]
\includegraphics[width=0.45\linewidth]{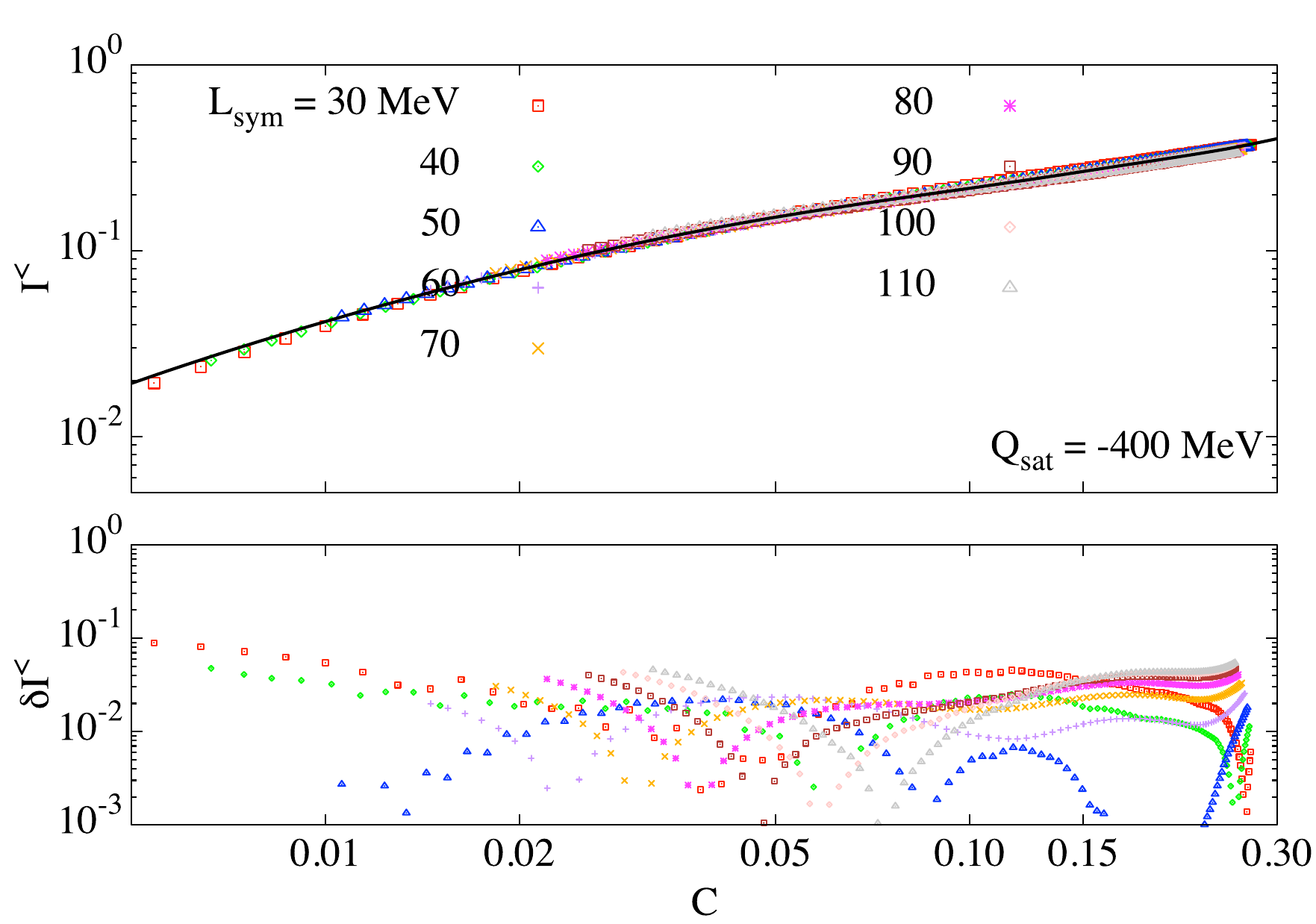}
\includegraphics[width=0.45\linewidth]{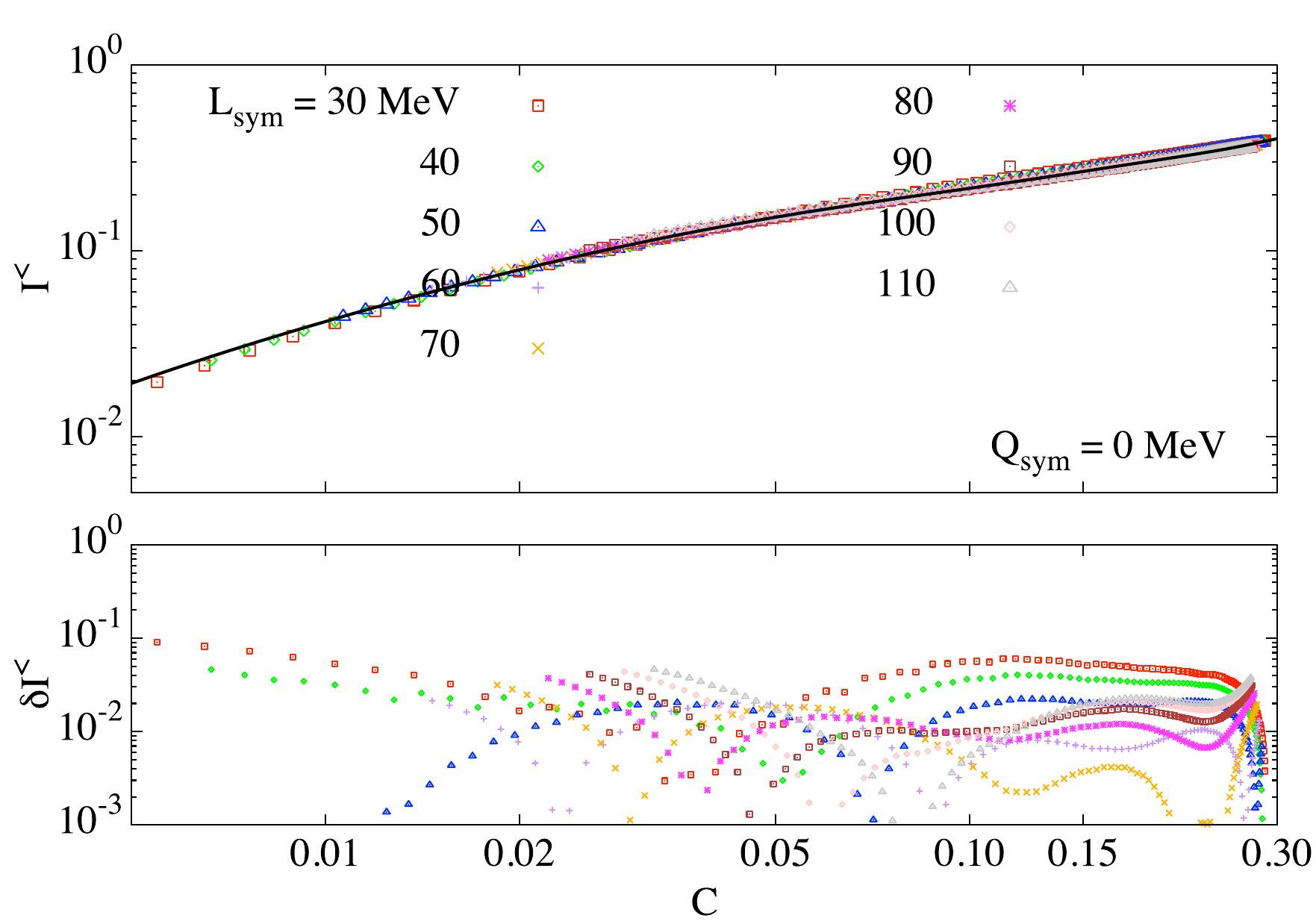}
\includegraphics[width=0.45\linewidth]{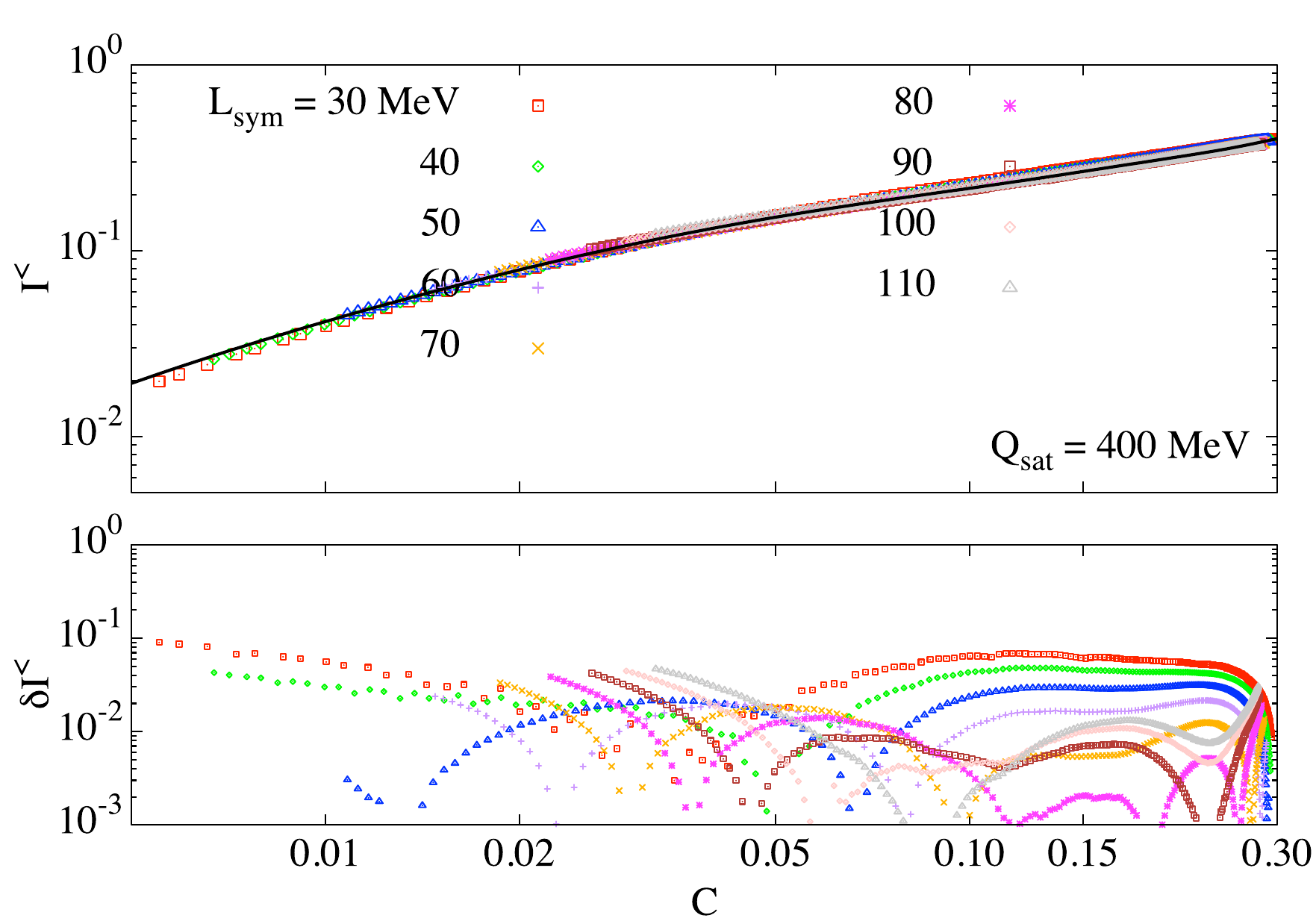}
\includegraphics[width=0.45\linewidth]{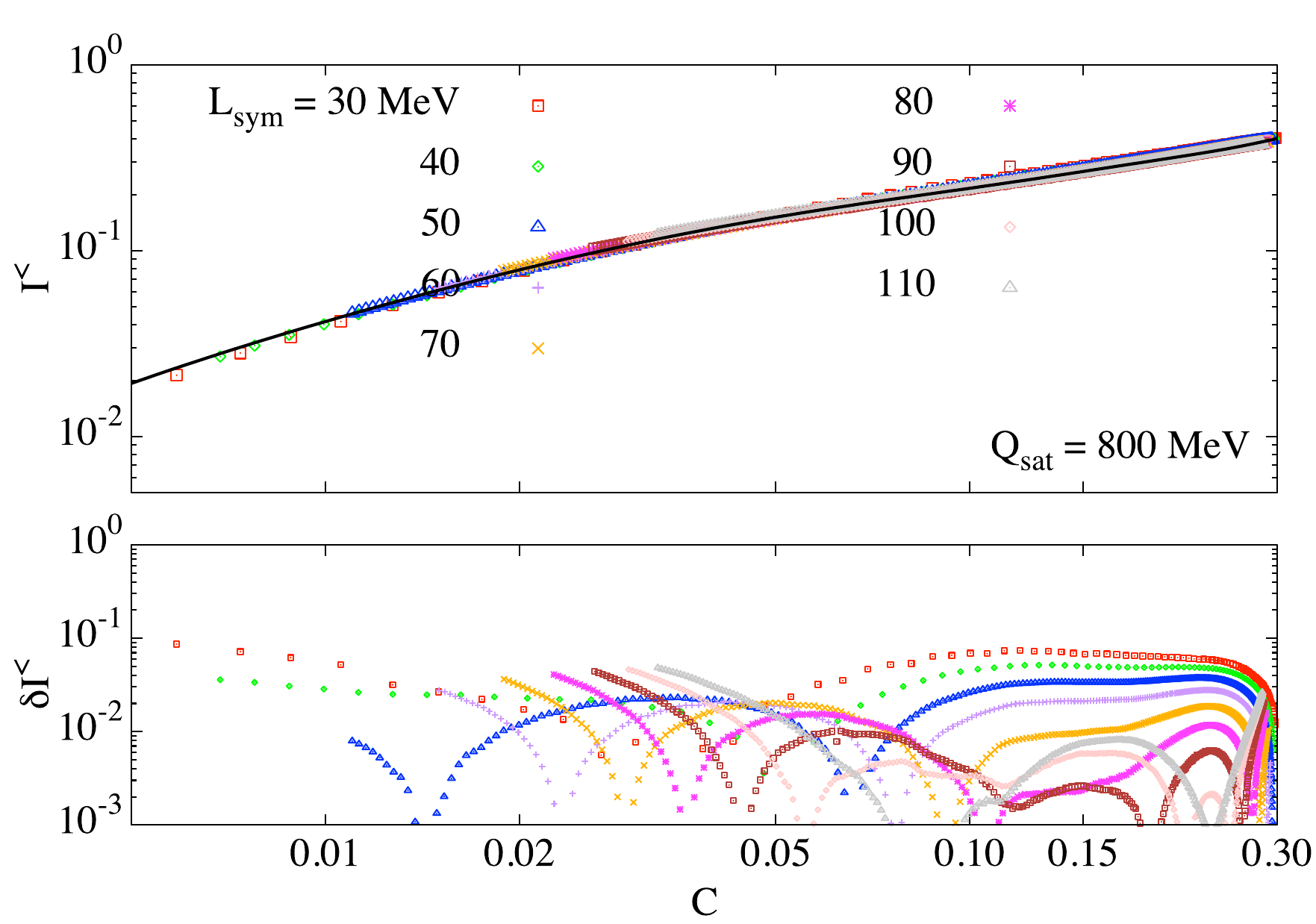}
\caption{Same as in Fig.~\ref{fig:I_bar_C_Fit_Error} but for $ I^{<} (C)$
  fitted using Eq.~\eqref{eq:I_mod} for maximally fast rotating
  (Keplerian) configurations.
}
    \label{fig:I_tilde_C_Fit_Error}
\end{figure*}

While it is well established that the macroscopic properties of CSs are highly sensitive to the underlying EoS, several universal relations have been identified between certain global quantities. Here, universality refers to remarkably weak dependence on the EoS, an observation that holds particularly well for cold, $\beta$-equilibrated CSs in the static limit. While the fundamental origin of these universal relations remains not fully understood, they have proven useful in practice. For example, they can be used to constrain otherwise inaccessible stellar parameters, reduce EoS-related uncertainties in data analysis, and break degeneracies between integral observables, such as those between the quadrupole moment and spin in gravitational waveforms from binary inspirals. In this section, we investigate the extent to which these universal relations persist in the case of cold, rapidly rotating CSs for a set of EoSs that exhibit continuous variations in the symmetry energy slope parameter $L_{\text{sym}}$ and isoscalar skewness coefficient $Q_{\text{sat}}$. Here, we focus on relations involving the normalized moment of inertia, the quadrupole moment, and the stellar compactness.

The theoretical minimum mass of a nonrotating neutron star lies in the range 0.2--0.9~$M_{\odot}$, depending on the EoS. However, neutron stars with masses $M \lesssim 1.17~M_{\odot}$ are generally expected to be unstable and unlikely to form via stellar collapse. Instead, such low-mass supernova remnants may result in the formation of a white dwarf.

Rapidly rotating neutron stars, particularly those near the Keplerian (mass-shedding) limit, can be stable at lower masses
compared with their nonrotating counterparts. Theoretical estimates place the minimum mass for such configurations in the range 0.1--0.5 $M_{\odot}$, again depending on the EoS. However, the astrophysical formation channels for these extremely low-mass, rapidly rotating neutron stars remain unclear. In the following, we consider sequences of Keplerian (mass-shedding limit) neutron star models with masses $M \geq 0.5\,M_{\odot}$. To properly fit the high-mass end of these sequences, we also include a small portion of the unstable branch (beyond the maximum mass) where the stellar models have masses slightly below the maximum (by a few percent).

We now turn to the two formulas that relate the moment of inertia to the compactness $C = M/ R$ where we define the compactness of rotating CSs using {\it the equatorial radius.} For nonrotating stars, the compactness varies in the range $0.1 \lesssim C \lesssim 0.3$. For rapidly rotating stars, both the mass and equatorial radius are increasing, but the equatorial radius increase overwhelms; therefore, the compactness of stars close to the maximum effectively decreases, compared with nonrotating neutron stars. 

The first fit formula was suggested in Ref.~\cite{Lattimer_ApJ_2005} for the normalized
moment of inertia ${I}^{<}=I /\left(MR^2\right)$ as a polynomial in compactness
\bea\label{eq:tilde_I} {I}^{<}(C)= \sum_{j=0}^m \alpha_j C^j, \eea
which was proven to be highly EoS insensitive. Reference~\cite{Breu:2016} revisited this problem, showing that an alternative universal relation using a different normalization of the moment of inertia, $I^> = I / M^3$, relates to compactness through inverse power expansion:
\bea\label{eq:bar_I} {I}^{>}(C)=\sum_{j=0}^m \alpha_j  [C^{j}]^{-1}.
\eea
\begin{table}
\begin{tabular}{|l|l|l|l|}
\hline \multicolumn{2}{|l|}{Final set of parameters} & \multicolumn{2}{|c|}{Asymptotic Standard Error} \\
\hline $\alpha_0 $ & $-2.97357 \times 10^{6}$ & $\pm 1.771\times 10^5$ & (5.957\%) \\
\hline $\alpha_1 $ & $-2.89435 \times 10^{6}$ & $\pm 1.723\times 10^5$ & (5.951\%) \\
\hline $\alpha_2 $ & $-1.36802 \times 10^{6}$ & $\pm 8.125\times 10^4$ & (5.939\%) \\
\hline $\alpha_3 $ & $ -40268 $ & $\pm 2.382\times 10^4$ & (5.915\%) \\
\hline $\alpha_4 $ &  $-76473.7 $ & $\pm 4414$ & (5.772\%) \\
\hline $\alpha_5 $ & $3.05279 \times 10^{6}$ & $\pm 1.82 \times 10^{5}$ & (5.962\%) \\
\hline $\alpha_6 $ &  $-79219.5$ & $\pm 4877$ & (6.156\%) \\
\hline
\end{tabular}
\caption{Parameters of the fit for the $I^<(C)$ function, Eq.~\eqref{eq:tilde_I}, after convergence is obtained, following 11 iterations with   $\sigma = 6.55706\times 10^{-3}$ and  $\chi_\nu^2= 4.2995\times 10^{-5}$.
}
\label{tab1}
\end{table}
\begin{table}
\begin{tabular}{|l|l|l|l|}
\hline \multicolumn{2}{|l|}{Final set of parameters} & \multicolumn{2}{|l|}{Asymptotic Standard Error} \\
\hline $\alpha_0$ & $ -6.46939$ & $\pm 0.04184$ & (0.6467\%) \\
\hline $\alpha_1$ & $2.62042$ & $\pm 0.006644$ & (0.2536\%) \\
\hline $\alpha_2$ & $0.0469605$ & $\pm 0.0002171$ & (0.4623\%) \\
\hline $\alpha_3$ & $-4.52455 \times 10^{-4}$ & $\pm2.33\times 10^{-6}$ & (0.515\%) \\
\hline $\alpha_4$ & $1.56228 \times 10^{-6}$ & $\pm 9.521 \times 10^{-9}$ & (0.6094\%) \\
\hline $\alpha_5$ & $-1.84774\times 10^{-9}$ & $\pm 1.293\times 10^{-11}$ & (0.6999\%) \\
\hline
\end{tabular}
\caption{Parameters of the fit for the $I^>(C)$ function, Eq.~\eqref{eq:bar_I},
  after convergence is obtained, following 13 iterations.
  The rms of residuals is $\sigma = 2.36535$ and the variance of residuals (reduced chi-square)
  $\chi_\nu^2= 5.59489$.  }
\label{tab2}
\end{table}

Equation \eqref{eq:bar_I} was fitted to the full data spanning $\Qsat$ and $\Lsym$ using $m=5$.
Equation  \eqref{eq:tilde_I} was modified by adding two additional exponential terms that improve
the fit in the low-$C\le 0.2$ regime, and the polynomial was truncated at $m=4$
\bea\label{eq:I_mod} {I}^{<}(C)= \sum_{j=0}^4 \alpha_j  C^{j}
+ \alpha_5 \exp(C) +\alpha_6 \exp(2C).
\eea
\begin{figure*}[tbh]
\includegraphics[width=0.45\linewidth]{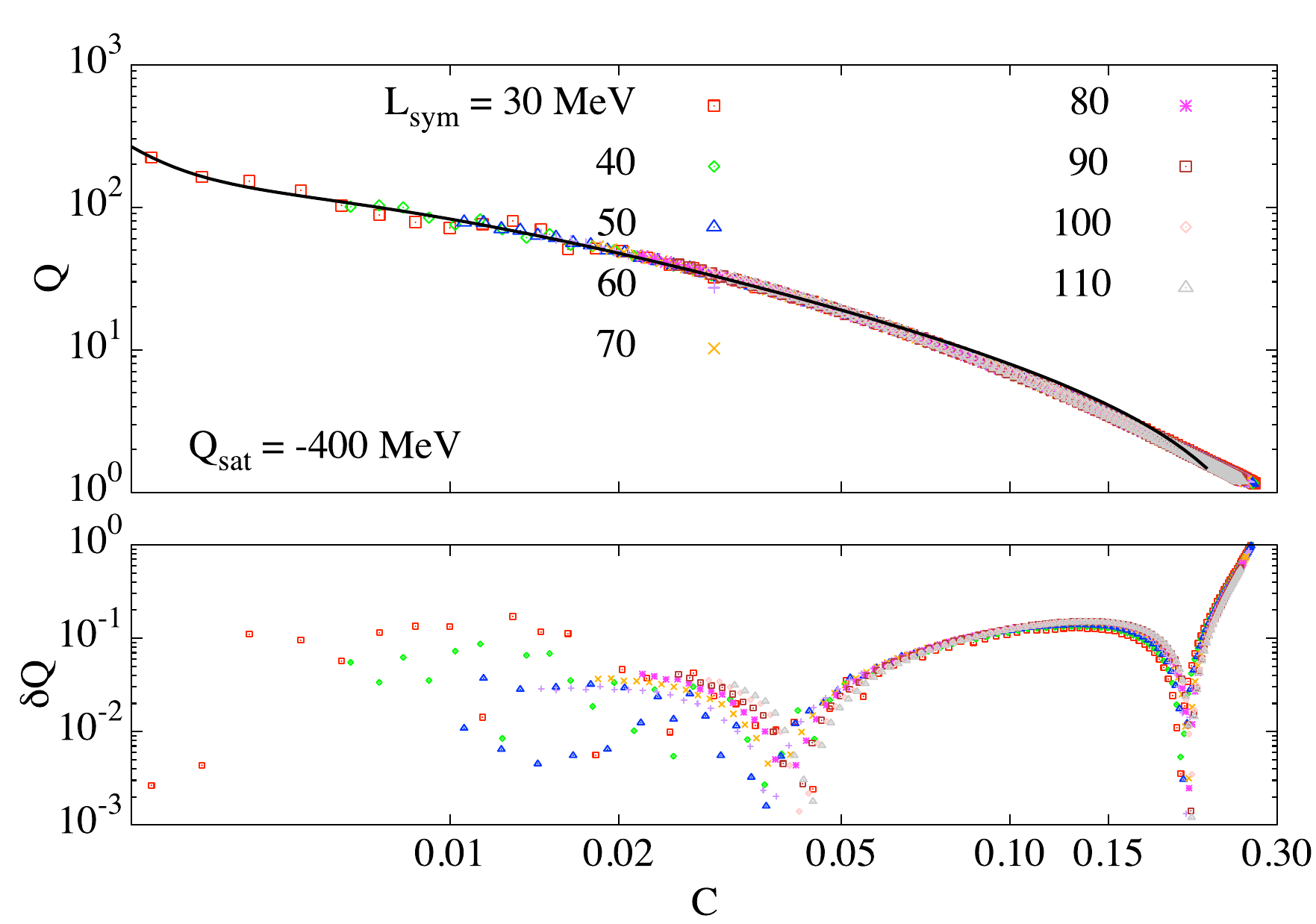}
\includegraphics[width=0.45\linewidth]{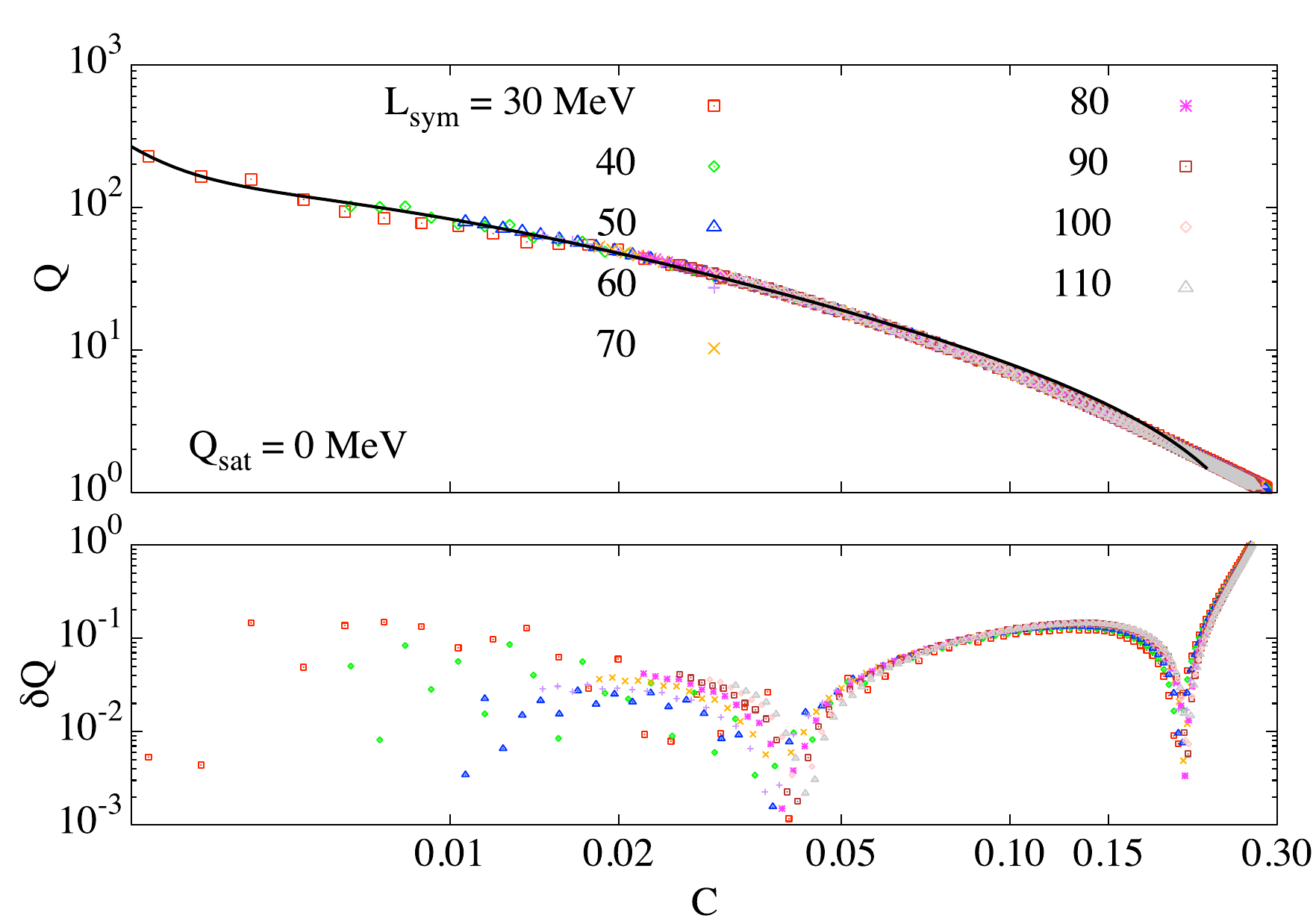}
\includegraphics[width=0.45\linewidth]{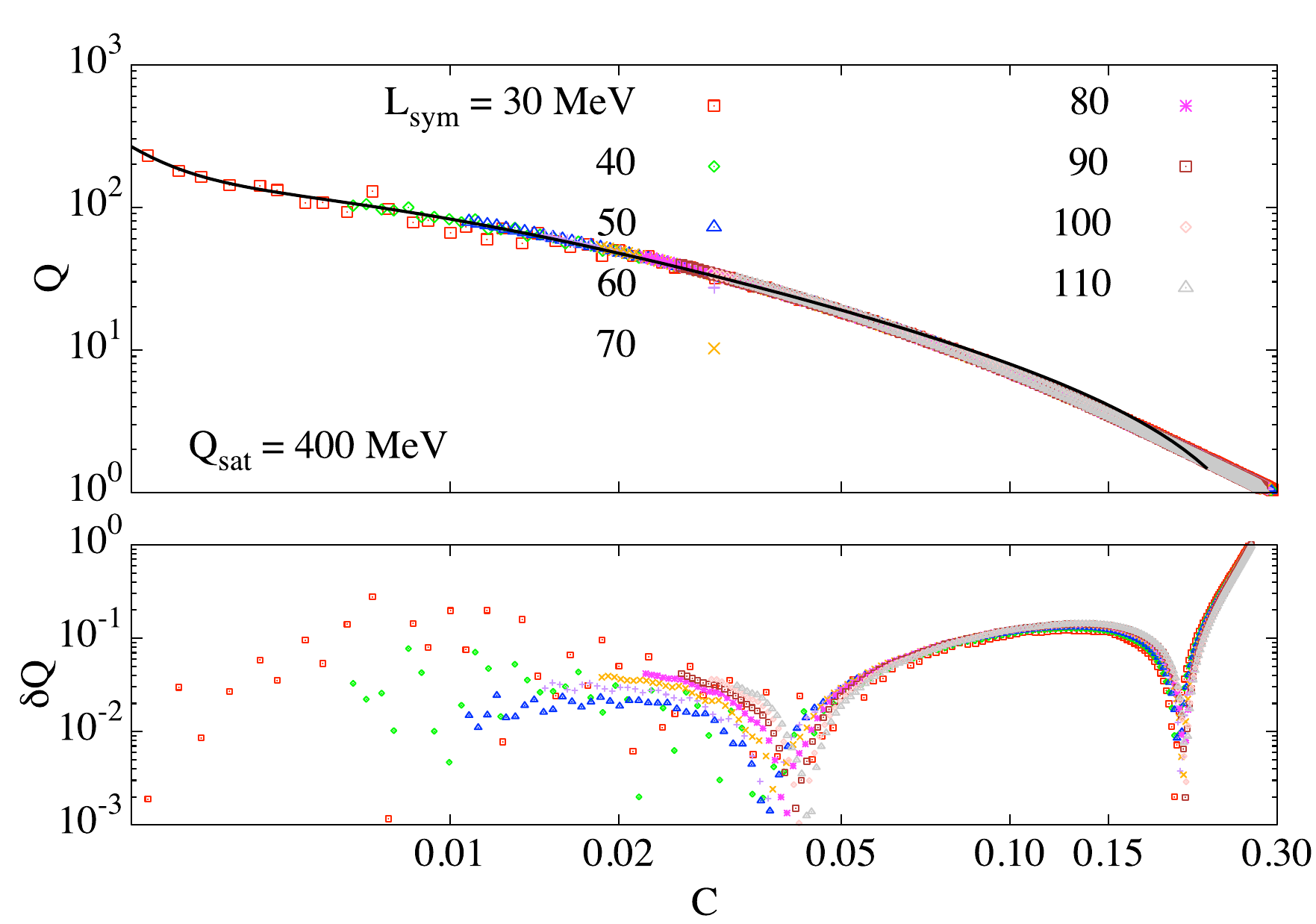}
\includegraphics[width=0.45\linewidth]{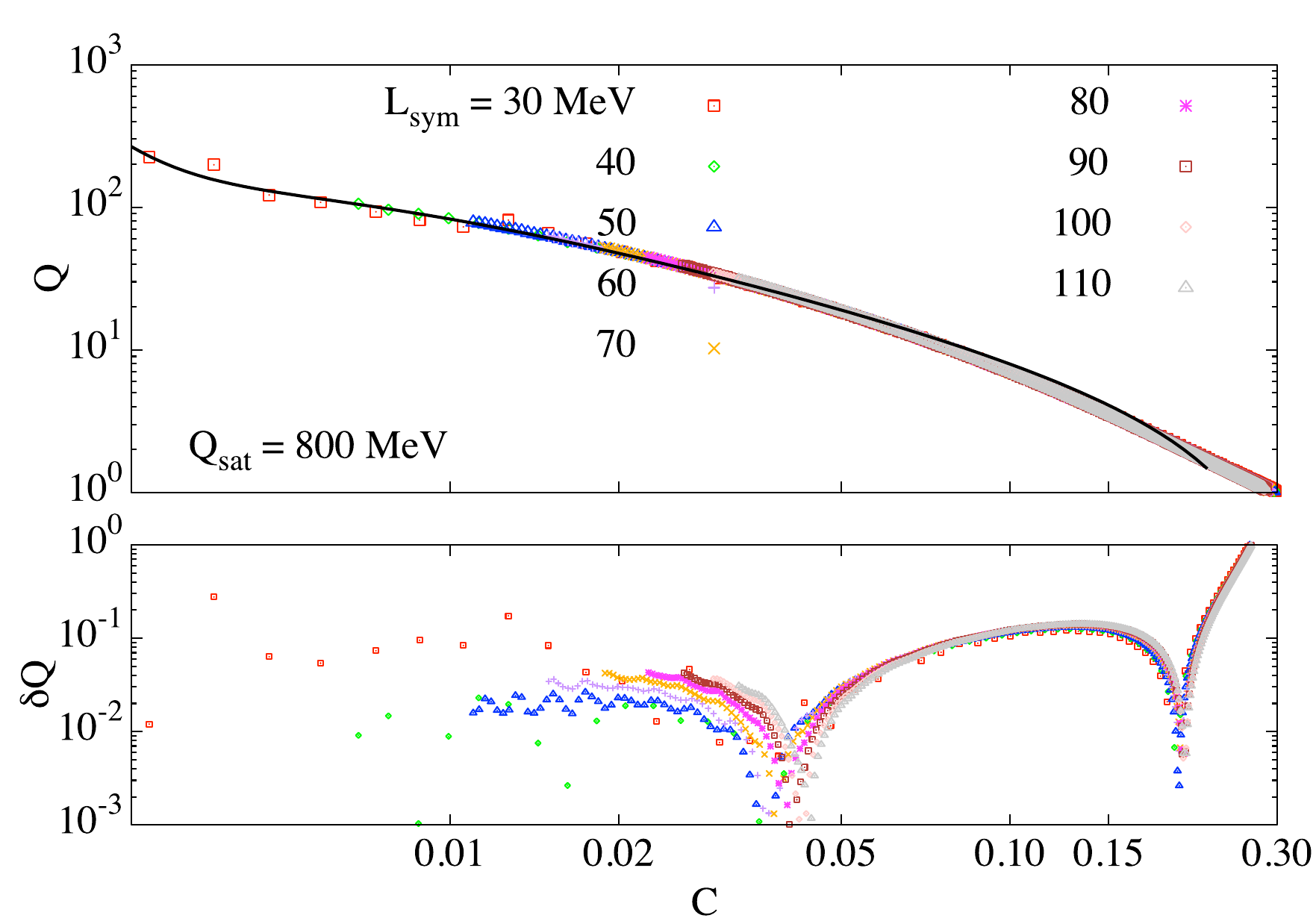}
\caption{Same as in Fig.~\ref{fig:I_bar_C_Fit_Error} but for the function $ Q(C)$ fitted
  using Eq.~\eqref{eq:bar_Q} for maximally fast rotating (Keplerian) configurations.
}
\label{fig:Q_bar_C_Fit_Error}
\end{figure*}
\begin{figure*}[tbh]
\includegraphics[width=0.45\linewidth]{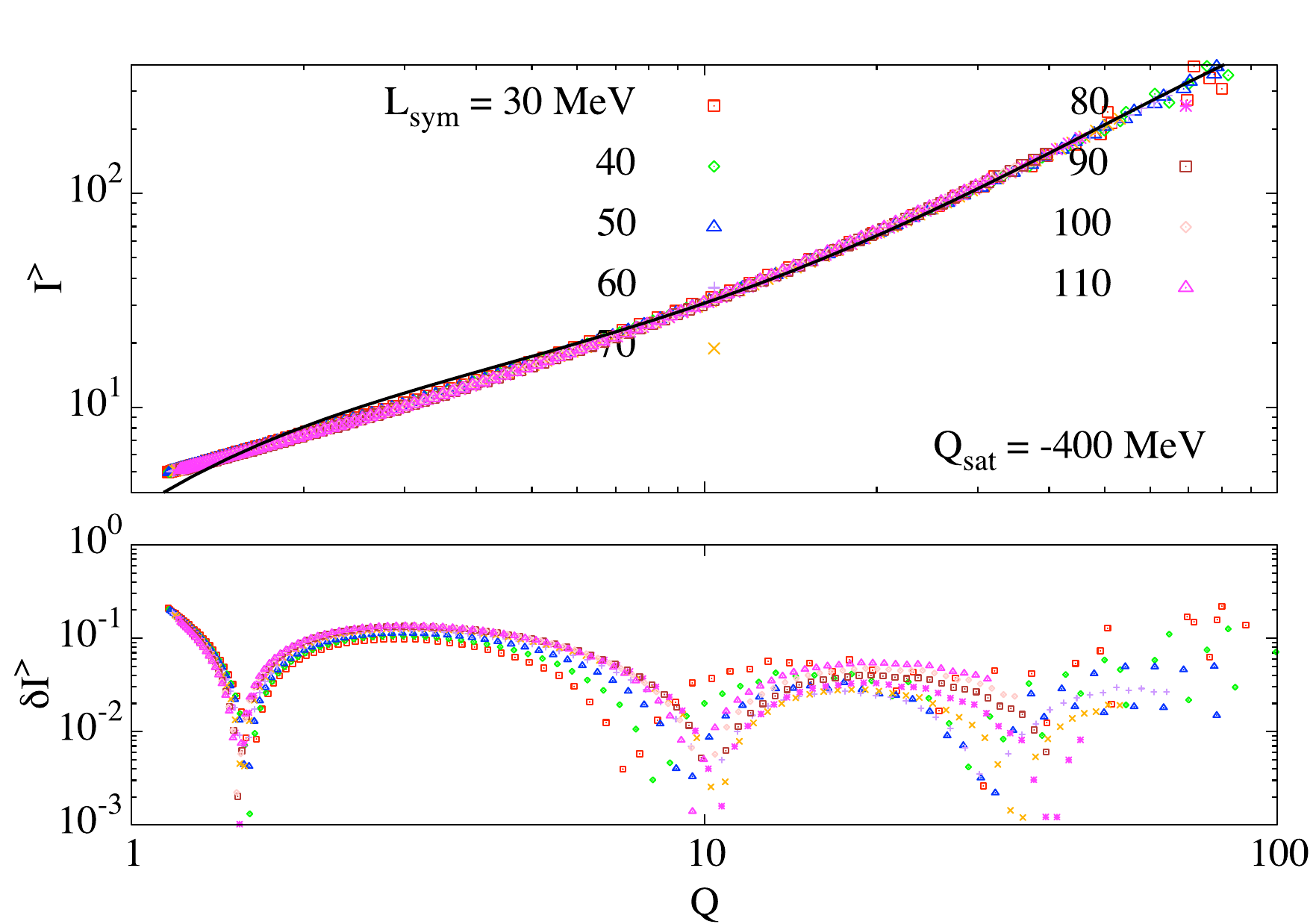}
\includegraphics[width=0.45\linewidth]{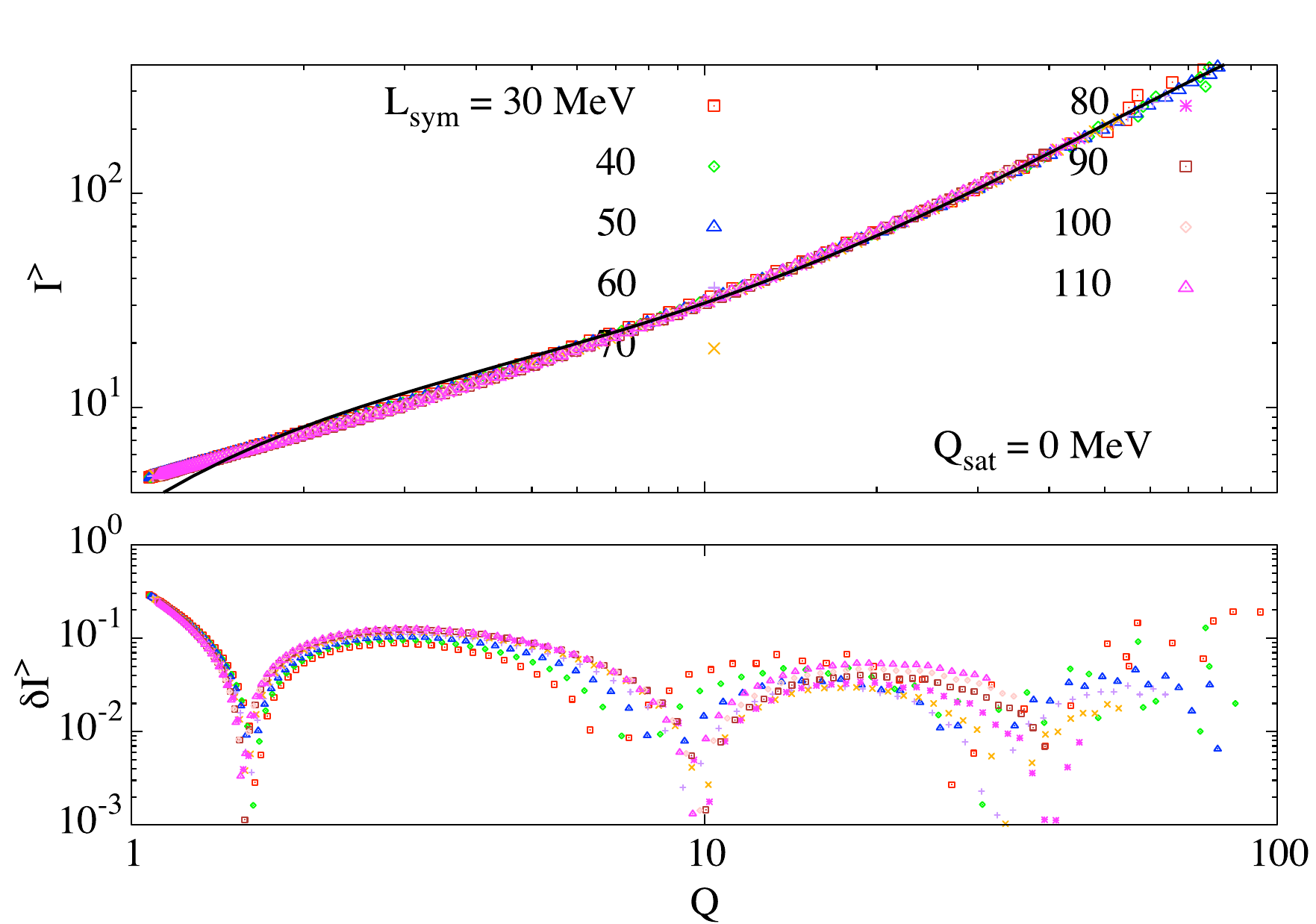}
\includegraphics[width=0.45\linewidth]{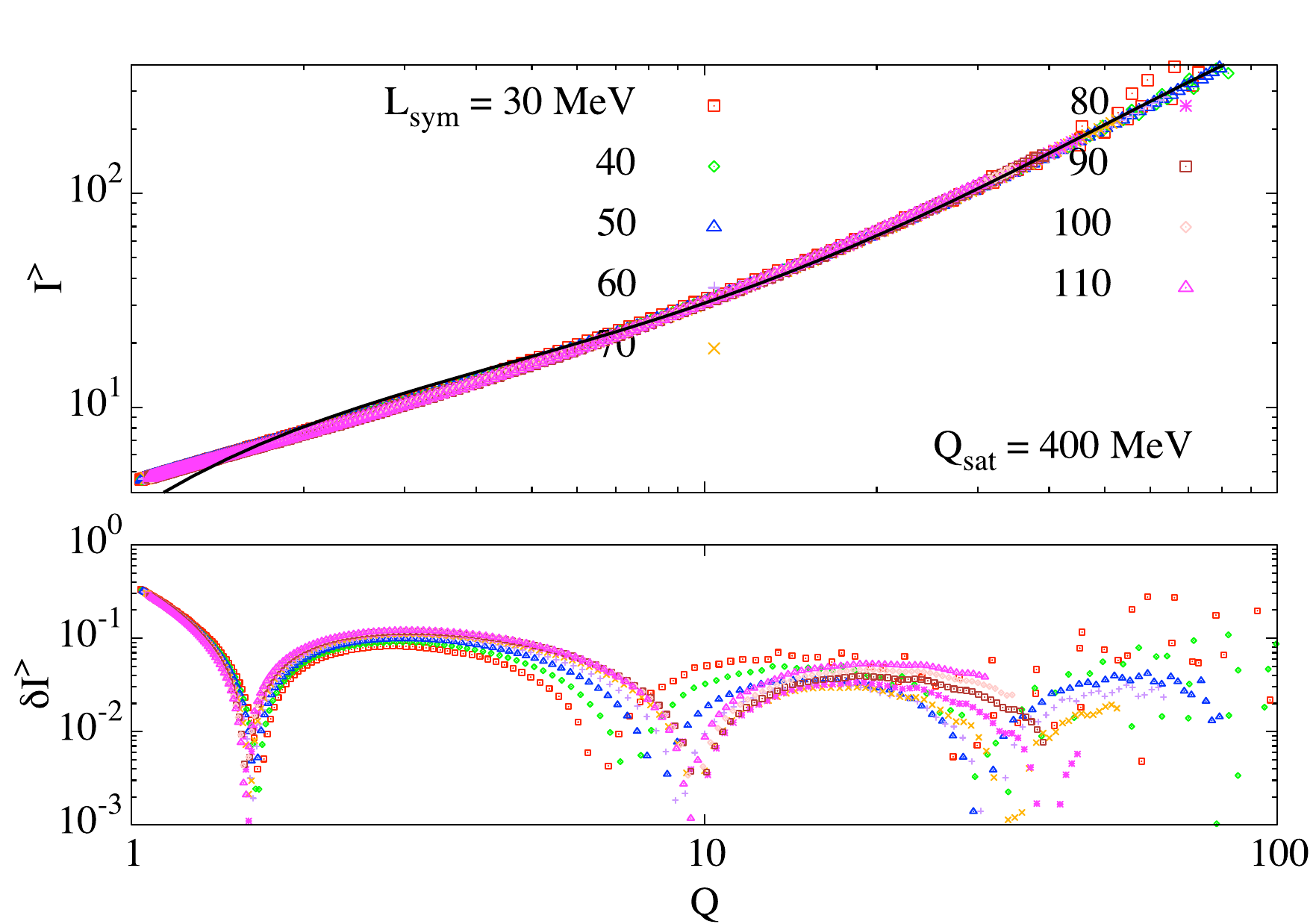}
\includegraphics[width=0.45\linewidth]{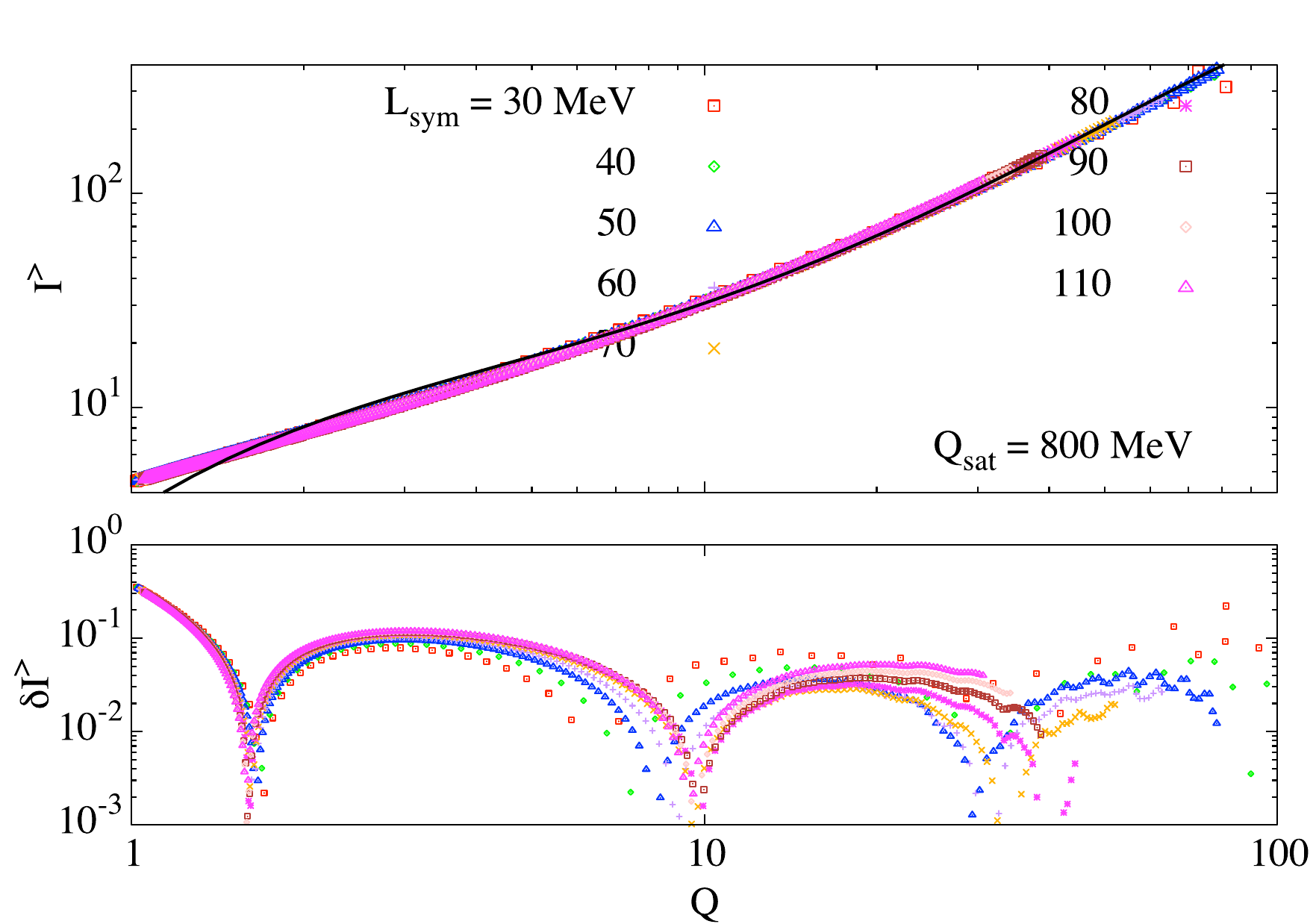}
\caption{Same as in Fig.~\ref{fig:I_bar_C_Fit_Error} but for the function $Q(I^>)$ fitted
  using Eq.~\eqref{eq:I_Q} for maximally fast rotating (Keplerian) configurations.
}
\label{fig:I_bar_Q_bar_Fit_Error}
\end{figure*}
The results are shown in Figs.~\ref{fig:I_bar_C_Fit_Error} and \ref{fig:I_tilde_C_Fit_Error}. The upper limit on $C$ is set by the maximum mass configuration.  In practice, we include stars beyond the maximum with masses that are below the maximum by a few percent.  The lower limit for $ C$ corresponds to stars in the mass range $0.1\le M/M_{\odot}\le 0.5$; the lower limit here corresponds to $L=30$~MeV while the upper limit to $L=110$~MeV. To enhance the clarity of the exposition, each panel in these figures shows the best fit performed on the full data, but only the data for fixed $\Qsat$ is shown in each panel. The results demonstrate that the proposed functional relations remain valid across the collection of EoSs under rapid rotation. Tables \ref{tab1} and \ref{tab2} list the values of the parameters for Eqs.~\eqref{eq:tilde_I} and \eqref{eq:bar_I}, respectively.

Reference~\cite{Yagi:2013a} demonstrated strong universal behavior in the relation between the normalized quadrupole moment and compactness. Defining
\bea\label{eq:Q_dimless} Q \equiv-\frac{Q^*}{M^3 \chi^2}
\eea
where $Q^*$ is the dimensional stellar (mass) quadrupole moment and $\chi=J / M^2$ is the dimensionless spin,  they found that $Q$ is nearly a polynomial function of the compactness, $C=M / R$, with a very weak dependence on the EoS, i.e., $Q(C)\simeq \sum_{j=0}^3 \alpha_jC^j,$ where the coefficients were empirically determined to high precision. Here we  use an alternative that relates the dimensionless quadrupole moment, Eq.~\eqref{eq:Q_dimless}, to the compactness via~\cite{Raduta_MNRAS_2020}
\bea\label{eq:bar_Q}
Q(C)\simeq \sum_{0}^m \alpha_j [C^{-j}].
\eea
Fits are performed using Eq.~\eqref{eq:bar_Q} with $m=4$ and the results are shown in
Fig.~\ref{fig:Q_bar_C_Fit_Error}. Table \ref{tab3} lists the values of parameters for Eq.~\eqref{eq:bar_Q}.
\begin{table}
\begin{tabular}{|l|l|l|l|}
\hline \multicolumn{2}{|l|}{Final set of parameters} & \multicolumn{2}{|c|}{Asymptotic Standard Error} \\
\hline $\alpha_0$ & $-3.90142$ &                   $\pm 0.01767 $                        &(0.453\%) \\
\hline $\alpha_1$ & $1.22695 $ &                      $\pm 0.00217$                        &(0.1769\%) \\
\hline $\alpha_2$ & $-4.21307\times 10^{-3}$ & $\pm  4.808 \times 10^{-5}$ & (1.141\%) \\
  \hline $\alpha_3$ & $5.29377 \times 10^{-5}$ & $\pm 2.992\times 10^{-7}$  & (5.652\%) \\
  \hline $\alpha_4$ & $6.60784 \times 10^{-9}$ & $\pm 5.337 \times 10^{-10}$ & (8.077\%) \\
  \hline
\end{tabular}
\caption{Parameters of the fit for the $Q(C)$ function, Eq.~\eqref{eq:bar_Q},  after convergence
  is obtained, following 7 iterations with 
 $\sigma = 1.20683$ and  $\chi_\nu^2= 1.45644$.}
\label{tab3}
\end{table}

Let us now examine the validity of the $I$-Love-$Q$ universal relations~\cite{Yagi:2013a,Yagi:2013b} for rapidly rotating CSs, focusing specifically on the ${I}^>-Q$ relation. For this purpose, we consider a collection of EoSs that feature continuous variation in the isoscalar skewness parameter $\Qsat$ and the symmetry energy slope parameter $\Lsym$. If the universality persists in this case, it is reasonable to expect that the remaining relations $I^>-\Lambda$ and $Q-\Lambda$ will also hold, given the strong correlation among these quantities observed in previous studies. Here $\Lambda=\lambda^{*}/ M^5$ is the dimensionless tidal deformability and $\lambda^{*}$ is its dimensional counterpart.

The corresponding fit formulas were established already in Refs.~\cite{Yagi:2013a,Yagi:2013b}, which used dimensionless moment of inertia $I^>$,  dimensionless quadrupole moment $Q$ [Eq.~\eqref{eq:Q_dimless}], and the dimensionless tidal deformability $\Lambda$.  These can be collectively written as an $m$th-order polynomial on a log-log scale
\bea\label{eq:I_Q}
\ln Y_i= \sum_{j=0}^{j=m} \alpha_j (\ln X_i)^j,
\eea
where $\left(Y_i, X_i\right)$ denotes one of the three quantity pairs listed above and $m=4$ in most cases. The coefficients $\alpha_j$ are fitted parameters that can be found from fits to data.  Although originally derived in the slow-rotation approximation, these universal relations have been shown to remain valid for rapidly rotating CSs, provided the fit parameters are adjusted to account for rotation frequency~\cite{Doneva:2013}. Given this evidence, we perform fits to the case where $Y_i\equiv I^>$ and $X_i \equiv  Q$ using Eq.~\eqref{eq:I_Q} in the Keplerian limit.

The corresponding data are obtained from the full collection of EoSs spanning all values $\Qsat$ and $\Lsym$. The fit results are shown in Fig.~\ref{fig:I_bar_Q_bar_Fit_Error}, where each panel compares the best fit to the data with a fixed value of $\Qsat$. It is seen that independent of the $\Qsat$ value, the results are uniformly accurate across the values of $\Lsym$. In this case, the deviations from the fitted curves are typically at
$\lesssim 10\%$. This is illustrated in the bottom panel of Fig.~\ref{fig:I_bar_Q_bar_Fit_Error}. We conclude that, for these quantities, universality is preserved for maximally rapidly rotating stars featuring different combinations of symmetry energy (represented by $\Lsym$) and high-density pressure in the symmetric matter (represented by $\Qsat$). The fit parameters are listed in Table~\ref{tab4}.
\begin{table}
\begin{tabular}{|l|l|l|l|}
  \hline \multicolumn{2}{|l|}{Final set of parameters} & \multicolumn{2}{|c|}{Asymptotic Standard Error} \\
  \hline $\alpha_0$ & $= 1.17189$    & $\pm 0.02158$     & (1.842\%) \\
 \hline $\alpha_1$  & $= 1.7559$       & $\pm 0.03648$    & (2.078\%) \\
 \hline $\alpha_2$  & $= -0.767783$ & $\pm 0.02085$    & (2.716\%) \\
\hline $\alpha_3$   & $= 0.237257$   & $\pm  0.004875$  & (2.055\%) \\
  \hline $\alpha_4$ & = -0.021993     & $\pm 0.0004255$ & (1.935\%) \\
  \hline
\end{tabular}
\caption{Parameters of the fit for the $I^>(Q)$ function according to \eqref{eq:I_Q} after convergence is obtained, following 13 iterations with 
  $\sigma = 5.94561$ and  $\chi_\nu^2= 35.3503$.}
\label{tab4}
\end{table}

\begin{figure*}[tbh]
\includegraphics[width=0.45\linewidth]{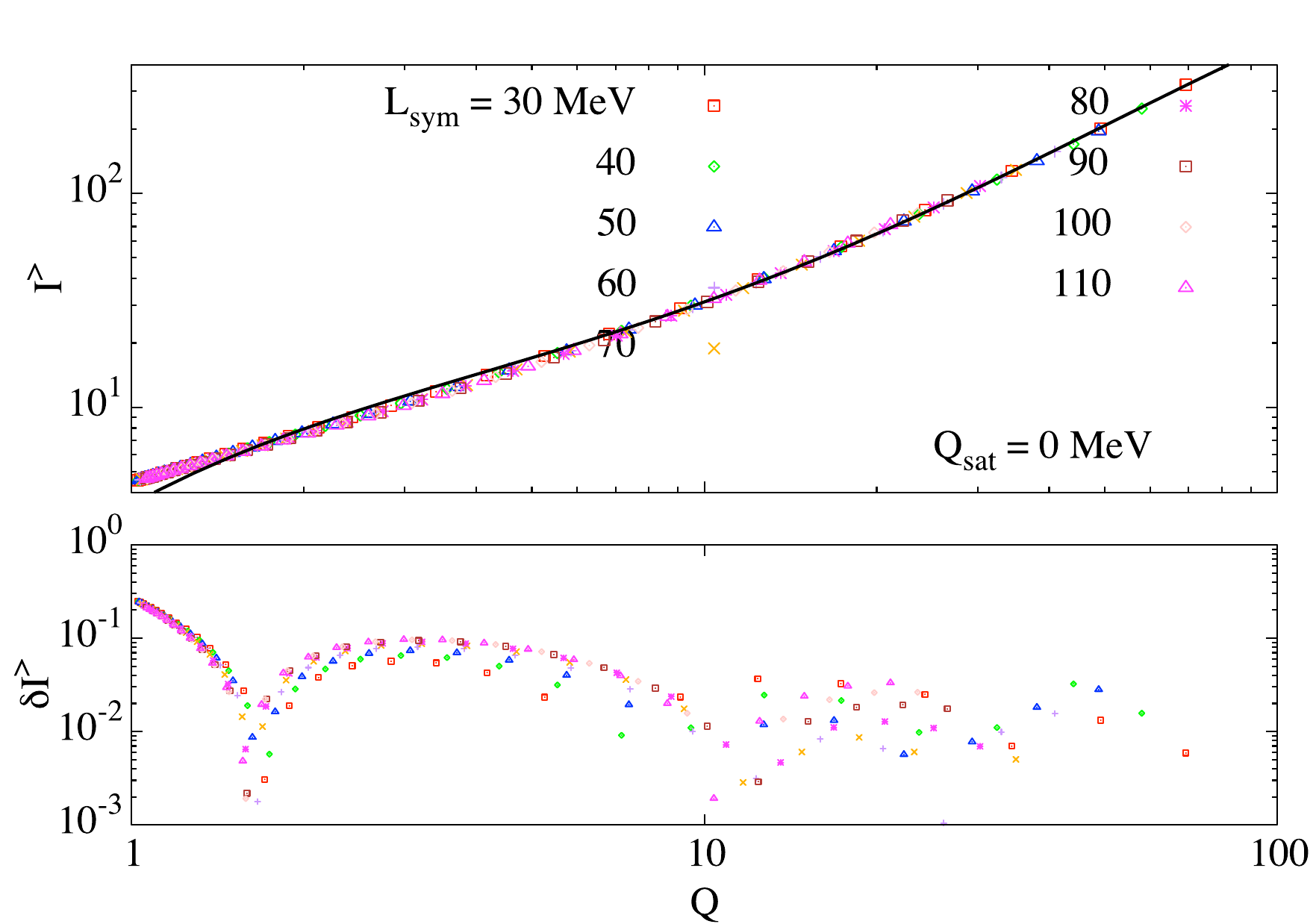}
\includegraphics[width=0.45\linewidth]{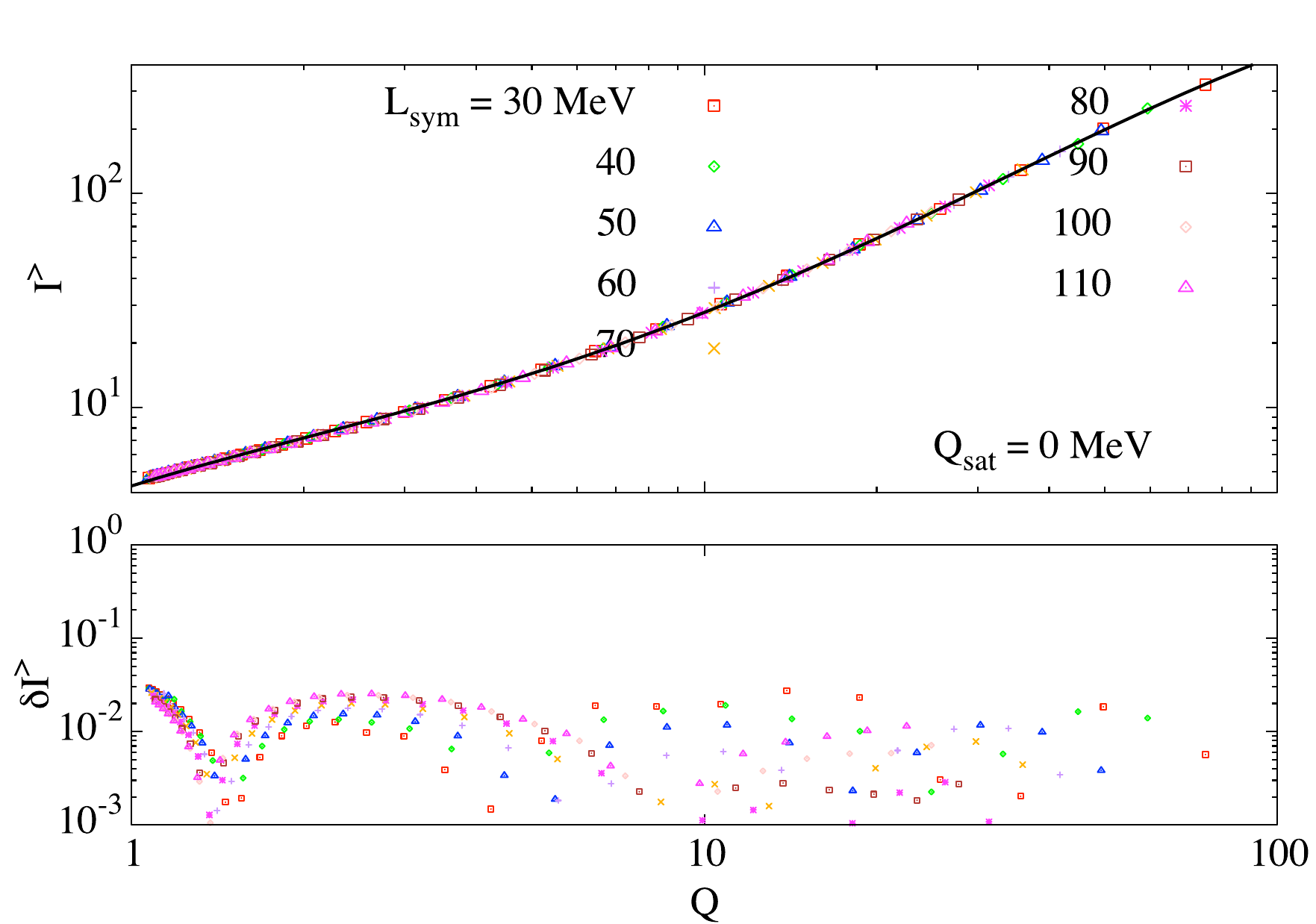}
\caption{Comparison of the function $Q(I^>)$ fitted
  using Eq.~\eqref{eq:I_Q} for Keplerian CSs (left panel) and CSs in the limit of slow rotation (right panel)
  The conventions are the same as in Fig.~\ref{fig:I_bar_C_Fit_Error}.
}
\label{fig:I_bar_Q_bar_Comparison}
\end{figure*}
  Given that the $I$-Love-$Q$ relations exhibit smaller errors in
  the slow-rotation limit than in the Keplerian limit, we 
  perform the computation in the slow-rotation regime by fixing the
  frequency to 1~Hz and using a reduced number of members (30) in each
  sequence defined by $\Qsat$ and $\Lsym$. The Keplerian and
  slow-rotation results are compared in
  Fig.~\ref{fig:I_bar_Q_bar_Comparison} for the case $\Qsat=0$. As
  expected, in the slow-rotation case, the relative error is
  suppressed to just a few percent, with the most significant
  differences appearing when $Q \leq 10$.

\section{Conclusions}
\label{sec:Conclusions}

In this work, we investigated universal relations for CSs rotating at
the Keplerian (mass-shedding) limit. Our analysis is based on the LS23
EoS set--a collection of nucleonic EoSs featuring systematic variations
in the symmetry energy slope parameter $\Lsym$ and the isoscalar
skewness parameter $\Qsat$. These parameters varied within ranges that
were broadly consistent with current experimental and astrophysical
constraints, though not always strictly confined to them.  We first
examined the global observable properties of isolated maximally
rotating stars, focusing on the mass-radius relation, moment of
inertia, quadrupole moment, and  dependence of the Keplerian
frequency on the stellar mass and $T/W$ ratio. Next, we
demonstrated that in the limit of Keplerian rotation, universal
relations remain valid across the set of EoSs characterized by varying
$\Lsym$ and $\Qsat$. In particular, we presented explicit results for
the moment of inertia and quadrupole moment as functions of
compactness, as well as for the $ I^>$--$ Q$ relation. All of
these relations display remarkable universality, with deviations
generally confined to a few percent and only rarely exceeding 10\%
over a broad range of parameters. In the slow-rotation limit, we
confirmed that the $I$–$Q$ relation in our set of equations attains
percent-level accuracy.  These findings support the applicability of
$I$–Love–$Q$-type universal relations in observational modeling of
maximally rotating CSs, even when accounting for significant variation
in the symmetry energy and high-density behavior of the nuclear EoS.

\section*{Acknowledgments}
This work was partly supported by the Polish National Science Centre (NCN) Grant 2023/51/B/ST9/02798 (A.~S. and S.~T.) and the Deutsche Forschungsgemeinschaft (DFG) Grant No. SE 1836/6-1 (A.~S). S.~T. is a member of the IMPRS for ``Quantum Dynamics and Control'' at the Max Planck Institute for the Physics of Complex Systems, Dresden, Germany, and acknowledges its partial support.  J.-J.~L. acknowledges the support of the National Natural Science Foundation of China under Grants No. 12105232 and No. 12475150.


\providecommand{\href}[2]{#2}\begingroup\raggedright\endgroup

\end{document}